\documentclass[twocolumn,trackchanges,resetfootnote]{aastex701}

\usepackage{comment}
\usepackage{multirow}
\usepackage[flushleft]{threeparttable}
\usepackage{graphicx}
\usepackage{amsmath}
\usepackage{breqn}
\shorttitle{Glimpse-D spectroscopy of faint-end galaxies}
\shortauthors{Asada et al.}
\graphicspath{{./}{figures/}}

\begin{document}

\title{GLIMPSE-DDT spectroscopic properties of faint-end galaxies at $z\sim6$:\\
Towards first metal enrichment, dust production, and ionizing photon production
}

\author[orcid=0000-0003-3983-5438,sname=Asada,gname=Yoshihisa]{Yoshihisa Asada}
\altaffiliation{Dunlap Fellow}
\affiliation{Dunlap Institute for Astronomy and Astrophysics, 50 St. George Street, Toronto, Ontario M5S 3H4, Canada}
\affiliation{Waseda Research Institute for Science and Engineering, Faculty of Science and Engineering, Waseda University,\\ 3-4-1 Okubo, Shinjuku, Tokyo 169-8555, Japan}
\email[show]{yoshi.asada@utoronto.ca}  

\author[0000-0001-7201-5066]{Seiji Fujimoto}
\affiliation{David A. Dunlap Department of Astronomy and Astrophysics, \\ University of Toronto, 50 St. George Street, Toronto, Ontario M5S 3H4, Canada}
\affiliation{Dunlap Institute for Astronomy and Astrophysics, 50 St. George Street, Toronto, Ontario M5S 3H4, Canada}
\email{seiji.fujimoto@utoronto.ca}

\author[0000-0002-0302-2577,sname=Chisholm,gname=John]{John Chisholm}
\affiliation{Department of Astronomy, The University of Texas at Austin, Austin, TX 78712, USA}
\affiliation{Cosmic Frontier Center, The University of Texas at Austin, Austin, TX 78712, USA}
\email{chisholm@austin.utexas.edu}

\author[0000-0003-3997-5705,sname=Naidu,gname=Rohan]{Rohan P.~Naidu}
\altaffiliation{NASA Hubble Fellow}
\affiliation{
MIT Kavli Institute for Astrophysics and Space Research, 70 Vassar Street, Cambridge, MA 02139, USA}
\email{rnaidu@mit.edu}

\author[0000-0002-7570-0824,sname=Atek,gname=Hakim]{Hakim Atek}
\affiliation{Institut d'Astrophysique de Paris, UMR 7095, CNRS, and Sorbonne Universit\'e, 98 bis boulevard Arago, 75014 Paris, France}
\email{hakim.atek@iap.fr}

\author[0000-0003-2680-005X,sname=Brammer,gname=Gabriel]{Gabriel Brammer}
\affiliation{Cosmic Dawn Center (DAWN), Niels Bohr Institute, University of Copenhagen, Jagtvej 128, K{\o}benhavn N, DK-2200, Denmark}
\email{gabriel.brammer@nbi.ku.dk}

\author[0000-0001-6278-032X]{Lukas J. Furtak}
\affiliation{Department of Astronomy, The University of Texas at Austin, Austin, TX 78712, USA}
\affiliation{Cosmic Frontier Center, The University of Texas at Austin, Austin, TX 78712, USA}
\email{furtak@utexas.edu}

\author[0000-0002-5588-9156,sname=Kokorev,gname=Vasily]{Vasily Kokorev}
\affiliation{Department of Astronomy, The University of Texas at Austin, Austin, TX 78712, USA}
\affiliation{Cosmic Frontier Center, The University of Texas at Austin, Austin, TX 78712, USA}
\email{vasily.kokorev.astro@gmail.com}

\author[0000-0002-9651-5716,sname=Pan,gname=Richard]{Richard Pan}
\affiliation{Department of Physics \& Astronomy, Tufts University, MA 02155, USA}
\email{richard.pan@tufts.edu}


\author[0000-0001-8104-9751]{Arghyadeep Basu}
\affiliation{Univ Lyon, Univ Lyon1, Ens de Lyon, CNRS, CRAL UMR5574, F-69230, Saint-Genis-Laval, France}
\email{arghyadeep.basu@univ-lyon1.fr}

\author[0000-0003-0212-2979]{Volker Bromm}
\affiliation{Weinberg Institute for Theoretical Physics, Texas Center for Cosmology and Astroparticle Physics, University of Texas at Austin, Austin, TX 78712, USA}
\affiliation{Department of Astronomy, The University of Texas at Austin, Austin, TX, USA}
\email{vbromm@astro.as.utexas.edu}

\author[0000-0002-5588-9156,sname=Dessauges-Zavadsky,gname=Miroslava]{Miroslava Dessauges-Zavadsky}
\affiliation{Department of Astronomy, University of Geneva, Chemin Pegasi 51, 1290 Versoix, Switzerland}
\email{miroslava.dessauges@unige.ch}

\author[0000-0003-4512-8705]{Tiger Yu-Yang Hsiao}\email{tiger.hsiao@utexas.edu}
\affiliation{Department of Astronomy, The University of Texas at Austin, Austin, TX 78712, USA}
\affiliation{Cosmic Frontier Center, The University of Texas at Austin, Austin, TX 78712, USA}
\email{tiger.hsiao@utexas.edu}

\author[0009-0004-4725-8559,sname=Jecmen,gname=Michelle]{Michelle Jecmen}
\affiliation{Department of Astronomy, The University of Texas at Austin, Austin, TX 78712, USA}
\email{mj32948@my.utexas.edu}

\author[0000-0002-3897-6856,gname=Damien,sname=Korber]{Damien Korber}
\affiliation{Department of Astronomy, University of Geneva, Chemin Pegasi 51, 1290 Versoix, Switzerland}
\email{damien.korber@unige.ch}

\author[0000-0002-4966-7450]{Boyuan Liu}
\affiliation{Institute of Astronomy, University of Cambridge, Madingley Road, Cambridge, CB3 0HA, UK}
\affiliation{Universit{\"a}t Heidelberg, Zentrum fur Astronomie, Institut f{\" u}r Theoretische Astrophysik, D-69120 Heidelberg, Germany}
\email{boyuan.liu.astro@gmail.com}

\author[0000-0002-6149-8178]{Jed McKinney}
\altaffiliation{NASA Hubble Fellow}
\affiliation{Department of Astronomy, The University of Texas at Austin, Austin, TX, USA}
\affiliation{Cosmic Frontier Center, The University of Texas at Austin, Austin, TX 78712}
\email{jed.mckinney.astro@gmail.com}

\author[0000-0001-5538-2614]{Kristen B. W. McQuinn}
\affiliation{Department of Physics and Astronomy, Rutgers University, Piscataway, NJ 08854, USA}
\affiliation{Space Telescope Science Institute, 3700 San Martin Drive, Baltimore, MD 21218, USA}
\email{kmcquinn@stsci.edu}

\author[0000-0001-7144-7182,gname=Dniel,sname=Schaerer]{Daniel Schaerer}
\affiliation{Department of Astronomy, University of Geneva, Chemin Pegasi 51, 1290 Versoix, Switzerland}
\affiliation{CNRS, IRAP, 14 Avenue E. Belin, 31400 Toulouse, France}
\email{daniel.schaerer@unige.ch}


\begin{abstract}
Ultra-faint galaxies at high-$z$ are fundamental elements of the early galaxy assembly, and spectroscopic characterization of this population is essential to understand the earliest galaxy evolution.
Leveraging the ultra-deep JWST/NIRCam and NIRSpec observations of a gravitational lensing field of Abell S1063, taken as part of the GLIMPSE survey, we present spectroscopic properties of 16 galaxies fainter than $M_{\rm UV}=-17$ mag, including the metallicity, dust attenuation, and the ionizing photon production efficiency.
The emission lines are generally quite strong, roughly half of which cannot be replicated with standard stellar populations and require an extreme ionizing source.
We also identify relatively strong [O{\sc iii}] emission lines from all sample galaxies, which indicates that the low-mass end of the mass-metallicity relation is extended down to $M_\star\sim10^6\ M_\odot$ at $z\sim6$. The strong [O{\sc iii}] line detection from the lowest-mass galaxy among the sample ($M_\star\sim10^{5.6}\ M_\odot$) stands in contrast to recent reports of extremely metal-poor galaxy candidates at similar mass and redshift, suggesting that there could be two distinct pathways of the earliest metal enrichment as simulations have predicted.
Interestingly, we detect both dust attenuation and galactic outflow in one of the sample galaxies with $M_\star=10^{6.6}\ M_\odot$ at $z=5.5$. All the dust, metal, and outflow contents in this galaxy can be consistently explained by supernovae (SNe), indicative of the key roles of SNe in the earliest galaxy assembly such as dust production, metal enrichment, stellar feedback, and baryon cycle.
\end{abstract}



\section{Introduction}\label{sec:intro}
Revealing the earliest phase of galaxy formation has been a key science driver in extragalactic astronomy.
Galaxies are first formed from the primordial environment via the Population III (Pop III) star-formation, and they eventually grow with acquiring stellar mass, generating heavy elements (metals) and producing dust via subsequent star-formation and stellar feedback.
Observationally characterizing this earliest evolutionary phase of galaxies is essential to understand how galaxies evolve from the initial conditions into mature galaxies observed in the present-day universe \citep[e.g.,][]{Bromm2011ARAA,Dayal2018PhR}.

With the advent of JWST, the detailed physical properties of relatively low-mass ($M_\star<10^9\ M_\odot$) galaxies at $z\gtrsim5$ have been probed with the exceptional sensitivity of NIRCam and NIRSpec up to $\lambda_{\rm obs}\sim5.5\ \mu$m \citep[e.g.,][]{Asada2024MNRAS,Endsley2024MNRAS,Harshan2024MNRAS,Atek2024Natur,Chemerynska2024ApJ,Topping2025MNRAS}.
Significant efforts have been made to look for lowest-metallicity galaxies at high-$z$ \citep[e.g.,][]{Willott2025ApJ,Hsiao2025arXiv,Nakajima2025arXiv,Morishita2025arXiv,Vanzella2025arXiv,Cai2025ApJ,Fujimoto2025ApJ}, some of which have successfully identified extremely metal-poor galaxy candidates (less than 1 \% solar metallicity) based on the lack of [O{\sc iii}]4959,5007 emission lines in relative to the H$\beta$ line.
Most of these metal-poor galaxy candidates are extremely faint (fainter than $M_{\rm UV}\sim-16$ mag) and low-mass (less than $M_\star\sim10^7\ M_\odot$).
The faintness of these systems make the detailed characterization of their interstellar medium (ISM) quite difficult even with JWST sensitivity in blank field surveys \cite[e.g.,][]{D'Eugenio2025ApJS}, and deep NIRSpec observations in gravitationally lensed field are necessary \citep[e.g.,][]{Morishita2025arXiv,Vanzella2025arXiv,Nakajima2025arXiv}.

So far, a number of $z>5$ sources with very high magnifications have been discovered with JWST \citep[e.g.,][]{Asada2023MNRAS,Adamo2024Natur,Bradac2025arXiv}.
The deep JWST/NIRSpec observations with high lens magnification enable us to study the ISM properties such as metallicity, dust, ionization parameters, or even kinematics in the lowest-mass galaxies at $z>5$ \citep[e.g.,][]{Mowla2024Natur,Fujimoto2025NatAs,Morishita2025arXiv,Vanzella2025arXiv}.
Such spectroscopic characterization of faint-end galaxies are also crucial to understand the reionization process of the universe at $z>6$ \citep[e.g.,][]{Atek2024Natur}, as they are often thought to be the main drivers of the Cosmic Reionization \citep[e.g.,][]{Robertson2015ApJ}.
Deep JWST/NIRSpec observations in gravitational lensing fields have successfully revealed the physical properties of low-mass end galaxies at $z>5$.

However, a statistical overview of their spectroscopic properties has been limited due to the inadequate sample size of highly magnified, faint-end $z>5$ galaxies with deep NIRSpec observations.
For example, \citet{Chemerynska2024ApJ} exploited NIRSpec/Prism observations of faint-end galaxies at $z\sim6-8$ behind the Abell 2744 cluster to study the mass-metallicity relation, but there are only 7 galaxies fainter than $M_{\rm UV}$ of $-17$ mag or less massive than $10^7\ M_\odot$.
Several case studies of extremely metal-poor galaxies have been presented \citep[e.g.,][]{Morishita2025arXiv,Vanzella2025arXiv}, while it is not clear if these galaxies are representative at their redshift/magnitude range.
It is thus essential to enlarge the sample size of high-$z$ faint-end galaxies with spectroscopic characterization, by leveraging deep NIRSpec observations in gravitational lensing fields, to push forward our understanding of the earliest galaxy evolution.

In this paper, we present a new sample of faint-end galaxies with gravitational lensing magnifications ($\langle\mu\rangle=2.16$), and provide an overview of their spectroscopic properties from deep NIRSpec observations.
We select our sample from the GLIMPSE survey \citep{Atek2025arXiv}, which conducted one of the deepest JWST NIRCam imaging observations in the gravitational lensing cluster AbellS 1063, and present spectroscopic follow-up observations taken as part of a JWST DDT program (PID:9223, PIs: S.~Fujimoto \& R.~Naidu) with NIRSpec in the micro-shutter array (MSA) mode.
The deep spectroscopic observations used the G395M/F290LP grating, with which we can explore the key ISM properties including metallicity, dust attenuation, ionizing photon productions, in faint, low-mass galaxies at $z>5$.
The $R\sim1000$ grating observation in a lensing field enables us not only to minimize the blending of key emission lines (e.g., H$\gamma$ and [O{\sc iii}]4363) but also to look for possible outflows from these faint-end galaxies.

The paper is structured as follows. 
In Sec.~\ref{sec:data}, a brief description of the data used and our data processing are given.
Sec.~\ref{sec:meas} outlines the sample selection and physical property measurements mainly based on the JWST NIRCam imaging and NIRSpec spectroscopy observations.
Sec.~\ref{sec:result} presents the main results, which includes the low-mass end of the mass-metallicity relation (Sec.~\ref{subsec:MZR}), the earliest dust production (Sec.~\ref{subsec:dust}), and production and escapes of ionizing photons in faint-end galaxies near the end of epoch of reionization (Sec.~\ref{subsec:xi_ion}).
In Sec.~\ref{sec:discussion}, using one galaxy in our sample, a comprehensive picture of the galaxy evolution regulated by supernovae is discussed.
The conclusion and summary is given in Sec.~\ref{sec:conclusion}.
Throughout the paper, a flat $\Lambda$-CDM cosmology with $\Omega_{m}=0.3$, $\Omega_{\Lambda}=0.7$, and $H_0=70\ {\rm km\ s^{-1}\ Mpc^{-1}}$ is assumed, and all magnitudes are quoted in the AB system \citep{Oke1983ApJ}.

\section{Data}\label{sec:data}
This work utilizes JWST/NIRCam imaging observations of the gravitational lensing cluster AbellS 1063 (AS1063, hereafter) by the GLIMPSE survey (PID:3293; PIs: H.~Atek \& J.~Chisholm; \citealt{Atek2025arXiv}), supplemented with HST/ACS+WFC3 observations taken as part of the Hubble Frontier Field program \citep{Lotz2017ApJL} and BUFFALO program \citep{Steinhardt2020ApJS}, and JWST/NIRSpec observations of the AS1063 field by a DDT program (PID:9223, PIs: S.~Fujimoto \& R.~Naidu).
The detailed survey description and NIRCam imaging data reduction by the GLIMPSE survey are described by \citet{Atek2025arXiv} and \citet{Kokorev2025arXiv}, and the NIRSpec observations and spectroscopic data reduction are detailed by \citet{Fei2025arXiv} and S.~Fujimoto et al. in prep.
We provide a brief description below.

\subsection{Imaging and photometry}
The GLIMPSE survey obtained NIRCam images in the AS1063 field with a total of 120 hours in seven broad-band filters and two medium-band filters; F090W, F115W, F150W, F200W, F277W, F356W, F410M, F444W, and F480M.
The NIRCam images are processed with the official JWST pipeline (version {\tt v12.0.9}) in conjunction with the Reference Data System (CRDS) context file {\tt jwst\_1321.pmap}, with a few minor tweaks to improve artefact removal. 
These include custom 1/f, local background and diffraction spike removal procedures, performed on an amplifier per amplifier basis. 
The NIRCam exposures are astrometrically aligned based on stars selected from Gaia.
The HST images are also reprocessed with the consistent astrometry alignment. JWST and HST exposures are then drizzled onto a common grid, with 0\farcs{02}/pix scale for NIRCam SW bands, and  0\farcs{04}/pix scale for NIRCam LW and HST mosaics. To improve source recovery, the bright cluster galaxies (bCGs) and the diffuse intra-cluster light (ICL) of the foreground galaxy cluster AS1063 were modeled and subtracted.
The bCGs are selected from the bCG catalog given by \citet{Shipley2018ApJS}, and model the bCG light profiles with \texttt{BMODEL} task in \texttt{IRAF/Isophote}.
We iterate the bCG modeling process to improve the image quality.
The final NIRCam mosaics achieve 5-sigma depth of $\sim30.8$ mag for a point source.

The photometric catalog is built leveraging the NIRCam images.
The source detection is performed on the noise-normalized NIRCam SW stack image with \texttt{SourceExtractor} \citep{Bertin1996AAPS}.
To recover LW-only sources, source detection is also ran on noise-normalized NIRCam LW stack image and sources that are not included in the SW-detection catalog are added to the final list of sources.
The source detection results in a total of 73621 sources.
For each detected source, photometry is performed using the \texttt{Photutils} package \citep{Bradley2020zndo}. 
The fixed-aperture photometry is performed on PSF-homogenized images varying the aperture size, ranging from $D=0.\!\!^{\prime\prime}1$ to $1.\!\!^{\prime\prime}2$.
The measured flux densities are then aperture-corrected based on the curve of growth of the F480M PSF.
Photometric uncertainties are estimated by placing 2000 empty apertures on the background sky area and adding the Poisson noise in quadrature.
In this work, we use the (aperture-corrected) flux measurement in $0.\!\!^{\prime\prime}2$-diameter apertures.

\subsection{NIRSpec data reduction}
The GLIMPSE-DDT program (PID: 9223; PIs: S.~Fujimoto \& R.~Naidu) conducted NIRSpec/MSA observation in the AS1063 field with the G395M/F290LP grating.
Three MSA configurations were adopted, each having $\sim$33-42 ks exposure times.
Some of the target were observed in multiple configurations, and a total of 384 sources were observed with the on-source exposure time varying from $\sim$9.2 hours to 30 hours.
The MSA spectra are processed with the \texttt{msaexp} package \citep[v0.9.8;][]{Brammer2022zndo} to do basic NIRSpec data reduction including $1/f$ noise correction, artefact removals, and bias correction for individual exposures.
The drizzle-combined 2D spectra are built after assigning WCS and flat fielding on each slitlet.
Then the background subtraction is performed locally based on the drizzle-combined 2D spectra in the source-free shutter.
The 1D spectra are extracted from the optimal aperture method by \citet{Horne1986PASP}.

\subsection{Gravitational magnification}
To determine the gravitational lensing magnifications, we use a new preliminary GLIMPSE SL model of AS1063 (to be published in Furtak et al.\ in prep., see also \citealt{Atek2025arXiv}), which is constructed with the new version of the \citet{Zitrin2015ApJ} parametric method, now updated to be fully analytic \citep[e.g.][]{Furtak2023MNRAS}. The mass distribution of the cluster is modeled with two components: a smooth dark matter (DM) component, parametrized with pseudo-isothermal elliptical mass distributions \citep[PIEMDs;][]{kassiola93}, and the cluster galaxies whose mass profiles are parametrized as dual pseudo-isothermal ellipsoids \citep[dPIE;][]{eliasdottir07}. Our lens model of AS1063 specifically comprises two cluster-scale DM halos and 303 cluster members. To constrain the lens model, we utilize spectroscopically confirmed multiple image compilations by \citet{Bergamini2019} and \citet{beauchesne2024}, and a handful of photometrically selected multiple images. In the end, the model is constrained with 75 multiple images of 28 sources, 24 of which have spectroscopic redshifts. The final model achieves an average lens plane image reproduction error $\Delta_{\mathrm{RMS}}=0.54\arcsec$.

Magnifications and uncertainties are computed analytically at each object's position and redshift. We take the median magnification value from the final MCMC chain as the fiducial magnification for each source, and estimated their uncertainties from their 16th- to 84th-percentiles.

\section{Measurements}\label{sec:meas}
\subsection{Sample selection}\label{subsec:selection}
\begin{figure}[t]
\centering
\includegraphics[width=0.9\linewidth]{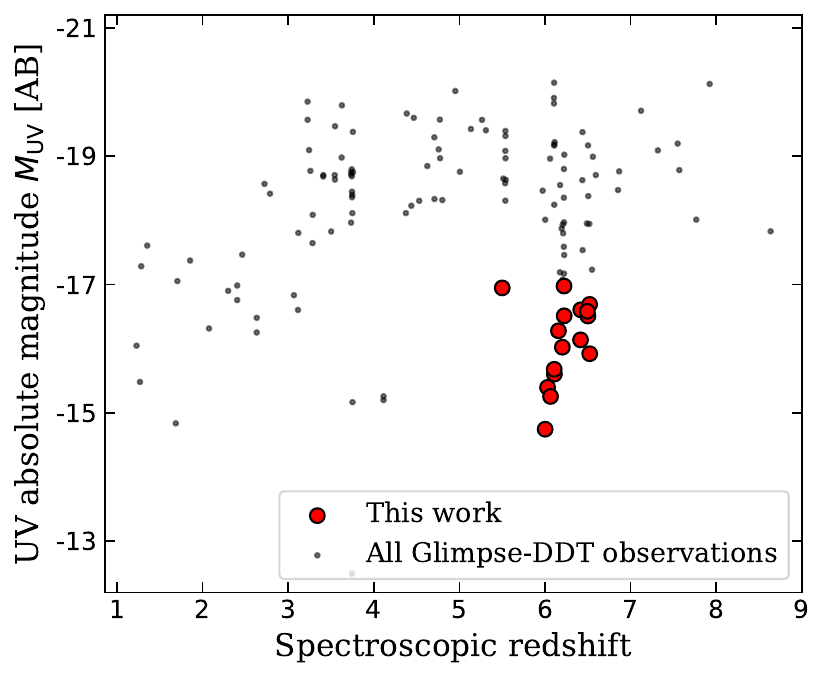}
\caption{Rest UV absolute magnitudes vs spectroscopic redshifts of the 16 sample galaxies (red filled circles). Black small circles present the entire galaxies targeted in the GLIMPSE-DDT observations.
The selected galaxies are fainter than $M_{\rm UV}=-17$ mag at $z_{\rm spec}>5$.}
\label{fig:Muv_zspec}
\end{figure}

We present a sample of UV-faint galaxies defined as fainter than $M_{\rm UV}=-17$ mag after lens correction.
We build the sample by requiring:
\begin{enumerate}
    \item clean NIRSpec spectrum without heavy contamination,
    \item reliable redshift measurements from one or more emission lines,
    \item NIRCam imaging coverage,
    \item $M_{\rm UV}>-17.0$ mag and $z_{\rm spec}>5$.
\end{enumerate}
The spectroscopic redshift is first measured with \texttt{msaexp} \citep{Brammer2022zndo} and the initial measurement is used in the sample selection.
The UV absolute magnitudes ($M_{\rm UV}$) are estimated from NIRCam photometry given the spectroscopic redshifts, correcting for the lens magnifications.
We obtain a sample of 16 UV-faint galaxies at $z_{\rm spec}>5$.
For the selected galaxies, we remeasure the spectroscopic redshifts by fitting a Gaussian to the H$\alpha$ emission line.
The selection yields a sample of 16 galaxies ranging from $z=5.5$ to $z=6.5$ and down to $M_{\rm UV}=-14.7$ mag.
Figure~\ref{fig:Muv_zspec} shows the distribution of the sample galaxies in the $M_{\rm UV}$ vs $z_{\rm spec}$ plane, compared to the entire NIRSpec targets of the GLIMPSE-DDT observation.

\subsection{Physical properties}\label{subsec:prop}
For each of the selected galaxies, we measure the basic physical properties from HST+NIRCam imaging and NIRSpec spectroscopic observations.
We first model the emission lines of H$\alpha$, H$\beta$, and [O{\sc iii}]4959/5007 with Gaussian profiles to obtain their line fluxes.
The continuum around the lines are simultaneously models as a constant offset.
The full list of emission line flux measurements are given in Appendix \ref{apx:line_fluxes}.
The H$\alpha$/H$\beta$ line ratios of the sample galaxies are consistent with expected value for the case B recombination (2.86) except for one galaxy ($3.01\pm0.05$ in GLIMPSE-55241; Sec.~\ref{subsec:dust}). We therefore apply the dust correction only for GLIMPSE-555241, assuming the Calzetti dust extinction curve \citep{Calzetti2000ApJ}.

These line fluxes are used to compute the star formation rate ($\textrm{SFR}_{\textrm{H}\alpha}$), ionizing photon production efficiency ($\xi_{\rm ion, obs}$), and oxygen abundance from the $R3$=[O{\sc iii}]5007/H$\beta$ line ratio.
The $\textrm{SFR}_{\textrm{H}\alpha}$ is converted from the H$\alpha$ luminosity assuming 10 \%\ solar stellar metallicity and accounting for the effect of binary stellar population \citep{Eldridge2017PASA}.
The ionizing photon production efficiency is estimated assuming the Case B recombination and taking the ratio to the rest UV luminosity measured from NIRCam photometry.
Note here we assume zero ionizing photon escape, and we call this dust-corrected but not $f_{\rm esc}$-corrected value as $\xi_{\rm ion, obs}$ distinguishing from the intrinsic values.
Both of $\textrm{SFR}_{\textrm{H}\alpha}$ and $\xi_{\rm ion, obs}$ are computed by scaling the NIRSpec spectrum to match the NIRCam photometry (i.e., the total flux measurements).
The oxygen abundance (12+$\log$(O/H)) is estimated from $R3$ based on the \citet{Nakajima2022ApJS} calibration for high EW sources in their work, since the $R3$-metallicity relation by \citet{Nakajima2022ApJS} is calibrated down to $R\sim1$, reached by our sample galaxies.
Finally, the gravitational lensing magnifications are corrected when needed, and the uncertainties in lens magnifications are propagated.
These measurements for each galaxy are listed in Table \ref{tab:sample}. Note that the errors on these measurements do not account for the systematic uncertainty due to the calibrations.
\citet{Nakajima2022ApJS} reported that their $R3$-metallicity relation has an intrinsic scatter of $\Delta\log({\rm O/H})=0.30$, thus the metallicity obtained in this work should also has a systematic uncertainty at least of 0.3 dex.

We next measure their stellar masses by spectro-photometry fitting using \texttt{Bagpipes} \citep{Carnall2019MNRAS} with \texttt{BPASS} \citep{Eldridge2017PASA} implementation.
We assume the Chabrier IMF \citep{Chabrier2003PASP} with the high-mass cut of 300 $M_\odot$, and we scale the NIRSpec spectra to match the NIRCam photometry during the fit.
We adopt the delayed-tau parametric star formation history and Calzetti dust extinction law \citep{Calzetti2000ApJ}, following the assumption in \citet{Atek2024Natur,Chemerynska2024ApJ}.
We also allow the redshift to vary with a Gaussian prior centered at the spectroscopic redshift.
In the end, we have eight free parameters in the spectro-photometry fitting (stellar mass $M_\star$, age, SFH $e$-holding time $\tau$, metallicity, dust attenuation, ionization parameter $U_{\rm ion}$, redshift, and the emission line width).
The resulting stellar masses of the sample range from $10^{5.6}\ M_\odot$ to $10^{7.0}\ M_\odot$.
The resulting posterior distributions and best estimate template spectra are shown in Appendix \ref{apx:sed_example}.

Table \ref{tab:sample} lists the resulting physical property measurements.
Their uncertainties are estimated from measurement errors and lens magnification uncertainties.
The stellar mass uncertainty from the SED fitting is based on the 16th-84th percentile range of the marginalized posterior.
For the H$\alpha$-based SFR, metallicity, and $\xi_{\rm ion}$, the uncertainties are obtained by propagating the line flux errors.
These uncertainties could thus be underestimated due to unknown systematics, including the intrinsic scatter of the $R3$-metallicity calibration, the choice of star-formation history, or the Case B assumptions entering $\xi_{\rm ion}$.

To explore the effect of SED fitting assumptions on stellar mass estimates, we also estimate the stellar masses with \texttt{Prospector}.
We use a non-parametric SFH approach with the ``\texttt{continuity\_sfh}'' prior by \citet{Leja2019ApJ}.
The model has six age bins, which are put at 0-10 Myr, 10-30 Myr, and the other four are logarithmically spaced up to $z=20$.
In Prospector fit, we adopt the Kroupa IMF \citep{Kroupa2002Sci} with the high-mass cut of $120 M_\odot$.
We find that the stellar mass estimates do not systematically change but can scatter by $\sim0.24$ dex when we change the fitting configuration.
We thus infer that the stellar masses can have systematic uncertainties of $\sim0.2-0.3$ dex but the main results in this paper are unchanged when these are taken into account.

\begin{longrotatetable}
\begin{deluxetable*}{lcccccccccc}
    \label{tab:sample}
    \tablecaption{Physical properties of the UV faintest galaxies from GLIMPSE-DDT observations.
  	}
    \tablewidth{0pt}
    \tablehead{
    \colhead{ID} & \colhead{R.A.} & \colhead{Decl.} & \colhead{$z_{\rm spec}$} & \colhead{$M_{\rm UV}$} & \colhead{$\beta_{\rm UV}$} & \colhead{$\log(M_\star/M_\odot)$} & \colhead{$\log$(SFR$_{\textrm{H}\alpha}$)} & \colhead{12+log(O/H)} & \colhead{$\xi_{\rm ion, obs}$} & \colhead{$\mu$} \\
    \colhead{} & \colhead{deg}& \colhead{deg} & \colhead{} & \colhead{mag} & \colhead{} & \colhead{} & \colhead{$M_\odot\ {\rm yr}^{-1}$} & \colhead{} &\colhead{Hz$^{-1}$ erg}\\
    \colhead{(1)} & \colhead{(2)}& \colhead{(3)} & \colhead{(4)} & \colhead{(5)} & \colhead{(6)} & \colhead{(7)} & \colhead{(8)} & \colhead{(9)} & \colhead{(10)} & \colhead{(11)} 
    }
    \startdata
    7685 & 342.246857 & -44.556080 & $6.4184 \pm 0.0004$ & $-16.61 \pm 0.10$ & $-2.71 \pm 0.32$ & $6.79^{+0.07}_{-0.07}$ & $-0.31 \pm 0.07$ & $7.30^{+0.14}_{-0.30}$ & $10^{25.94 \pm 0.08}$ & $1.26 \pm 0.02$ \\
8139 & 342.261292 & -44.555302 & $6.2233 \pm 0.0003$ & $-16.51 \pm 0.16$ & $-2.08 \pm 0.34$ & $6.80^{+0.08}_{-0.09}$ & $-0.33 \pm 0.04$ & $7.38^{+0.06}_{-0.13}$ & $10^{25.97 \pm 0.07}$ & $1.22 \pm 0.01$ \\
9081 & 342.256592 & -44.553780 & $6.2218 \pm 0.0002$ & $-16.98 \pm 0.08$ & $-2.28 \pm 0.19$ & $7.00^{+0.04}_{-0.04}$ & $-0.22 \pm 0.04$ & $7.71^{+0.35}_{-0.56}$ & $10^{25.89 \pm 0.05}$ & $1.24 \pm 0.02$ \\
11789 & 342.254211 & -44.549503 & $6.5210 \pm 0.0007$\tablenotemark{$\dagger$} & $-16.69 \pm 0.09$ & $-2.50 \pm 0.27$ & $6.84^{+0.06}_{-0.07}$ & $-0.04 \pm 0.08$\tablenotemark{$\dagger$} & $7.17^{+0.09}_{-0.18}$ & $10^{26.19 \pm 0.08}$\tablenotemark{$\dagger$} & $1.26 \pm 0.02$ \\
13186 & 342.173615 & -44.547592 & $6.1572 \pm 0.0002$ & $-16.28 \pm 0.09$ & $-2.17 \pm 0.21$ & $6.40^{+0.18}_{-0.16}$ & $-0.27 \pm 0.05$ & $6.94^{+0.08}_{-0.16}$ & $10^{26.12 \pm 0.05}$ & $3.43 \pm 0.15$ \\
17070 & 342.178253 & -44.543865 & $6.0297 \pm 0.0001$ & $-15.40 \pm 0.07$ & $-2.45 \pm 0.23$ & $6.34^{+0.06}_{-0.06}$ & $-0.68 \pm 0.05$ & $7.10^{+0.09}_{-0.19}$ & $10^{26.06 \pm 0.05}$ & $5.31 \pm 0.28$ \\
19034 & 342.212372 & -44.528755 & $6.2027 \pm 0.0002$ & $-16.02 \pm 0.11$ & $-2.08 \pm 0.27$ & $6.57^{+0.07}_{-0.06}$ & $-0.92 \pm 0.05$ & $6.98^{+0.05}_{-0.10}$ & $10^{25.57 \pm 0.06}$ & $2.79 \pm 0.15$ \\
35552 & 342.236603 & -44.533451 & $6.5018 \pm 0.0007$ & $-16.51 \pm 0.09$ & $-2.45 \pm 0.23$ & $6.95^{+0.06}_{-0.05}$ & $-0.78 \pm 0.09$ & $7.31^{+0.12}_{-0.25}$ & $10^{25.52 \pm 0.09}$ & $1.53 \pm 0.04$ \\
35957 & 342.183899 & -44.535336 & $6.1070 \pm 0.0001$ & $-15.68 \pm 0.03$ & $-2.61 \pm 0.12$ & $6.51^{+0.06}_{-0.06}$ & $-1.03 \pm 0.07$ & $7.49^{+0.10}_{-0.20}$ & $10^{25.60 \pm 0.05}$ & $16.38 \pm 1.80$ \\
37914 & 342.188446 & -44.536194 & $6.1072 \pm 0.0002$ & $-15.60 \pm 0.05$ & $-2.58 \pm 0.16$ & $6.47^{+0.07}_{-0.07}$ & $-0.98 \pm 0.08$ & $7.21^{+0.08}_{-0.17}$ & $10^{25.68 \pm 0.06}$ & $13.47 \pm 1.62$ \\
45084 & 342.258545 & -44.540390 & $6.4940 \pm 0.0003$ & $-16.58 \pm 0.09$ & $-2.46 \pm 0.24$ & $6.82^{+0.06}_{-0.06}$ & $-0.45 \pm 0.05$ & $7.11^{+0.08}_{-0.17}$ & $10^{25.81 \pm 0.06}$ & $1.30 \pm 0.02$ \\
45706 & 342.236816 & -44.540749 & $6.4153 \pm 0.0003$ & $-16.14 \pm 0.14$ & $-2.32 \pm 0.36$ & $6.16^{+0.08}_{-0.04}$ & $0.08 \pm 0.05$ & $7.00^{+0.07}_{-0.14}$ & $10^{26.52 \pm 0.07}$ & $1.42 \pm 0.03$ \\
46431 & 342.236511 & -44.541180 & $6.5224 \pm 0.0006$ & $-15.92 \pm 0.13$ & $-2.89 \pm 0.46$ & $6.28^{+0.20}_{-0.16}$ & $0.02 \pm 0.07$ & $7.09^{+0.14}_{-0.31}$ & $10^{26.55 \pm 0.08}$ & $1.43 \pm 0.02$ \\
47757 & 342.175934 & -44.542404 & $6.0019 \pm 0.0001$ & $-14.74 \pm 0.05$ & $-2.40 \pm 0.15$ & $5.62^{+0.06}_{-0.06}$ & $-1.00 \pm 0.05$ & $7.15^{+0.03}_{-0.06}$ & $10^{26.00 \pm 0.05}$ & $13.36 \pm 1.40$ \\
54637 & 342.205902 & -44.519085 & $6.0654 \pm 0.0006$ & $-15.26 \pm 0.04$ & $-2.67 \pm 0.11$ & $6.47^{+0.25}_{-0.17}$ & $-2.02 \pm 0.24$ & $7.22^{+0.12}_{-0.25}$ & $10^{24.78 \pm 0.17}$ & $18.91 \pm 8.09$ \\
55241 & 342.167999 & -44.542530 & $5.4996 \pm 0.0000$ & $-16.95 \pm 0.03$ & $-2.01 \pm 0.01$ & $6.76^{+0.09}_{-0.08}$ & $-0.33 \pm 0.08$ & $7.54^{+0.01}_{-0.02}$ & $10^{25.85 \pm 0.08}$ & $17.05 \pm 3.30$ \\
    \enddata
    \tablenotetext{}{(1) Source IDs in the GLIMPSE catalog. (2) Right Ascension in J2000. (3) Declination in J2000. (4) Spectroscopic redshifts. (5) Absolute rest UV magnitudes. (6) UV slopes measured from NIRCam photometry. (7) Stellar masses. (8) SFRs from the H$\alpha$ line luminosity. (9) Oxygen abundance from $R3$=[O{\sc iii}]5007/H$\beta$ line ratios. (10) Observed ionizing photon production efficiency. (11) Magnification factors.
    }
    \tablenotetext{}{$\dagger$ The H$\alpha$ line falls in the gap of NIRSpec detectors and the H$\beta$ line is used instead.}
\end{deluxetable*}
\end{longrotatetable}

\section{Results}\label{sec:result}
\subsection{Low-mass end of the Mass-Metallicity relation}\label{subsec:MZR}
\begin{figure*}[t]
\centering
\includegraphics[width=0.95\textwidth]{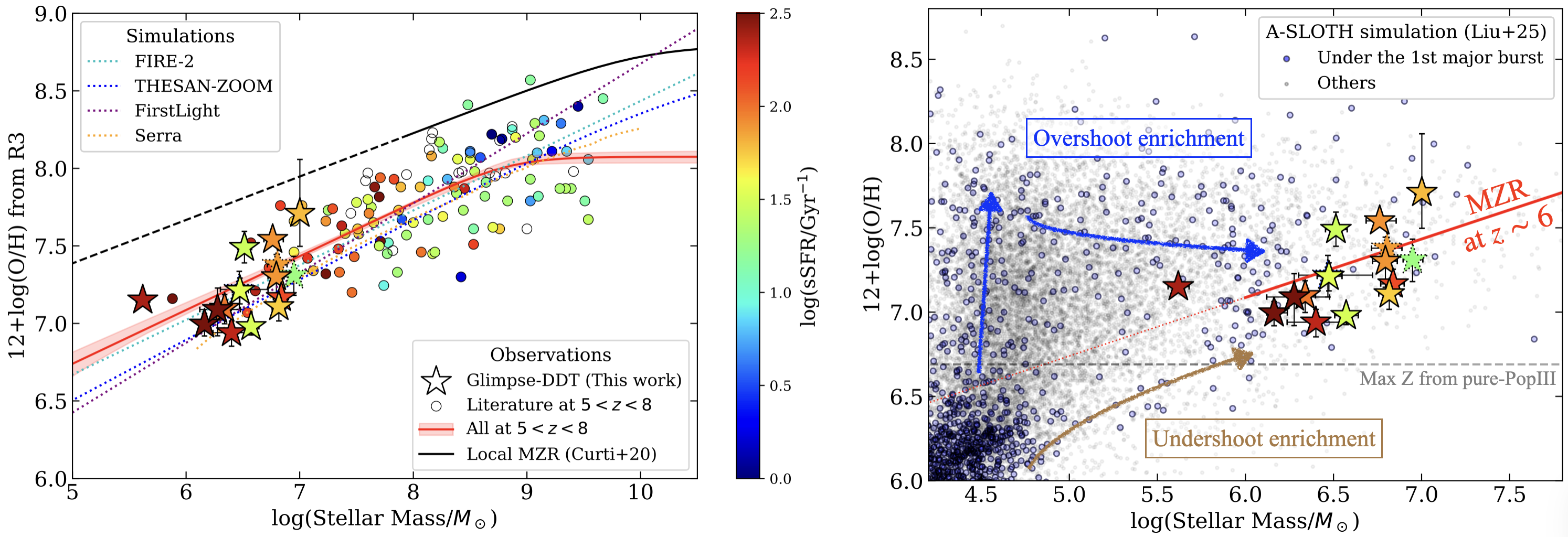}
\caption{The mass-metallicity relation (MZR) at $z\sim6$. \textit{Left}: the observed MZR in this work (stars) shown with measurements in literature \citep[circles;][]{Nakajima2023ApJS,Curti2024AAP,Chemerynska2024ApJ,Topping2025ApJ}, color-coded with the sSFR. 
Sources lacking either H$\alpha$ or H$\beta$ detection in our sample are shown by dashed stars.
The red solid curve is the best fit MZR parameterization (Eqn.~\ref{eqn:MZR}) at $5<z<8$, while the red shaded area denotes the 16th to 84th percentile range of the posterior. The black curve presents the local MZR, while colored dotted curves show MZR predictions from cosmological simulations (\textsc{FirstLight}, \citealt{Langan2020MNRAS}; \textsc{Thesan-Zoom}, \citealt{McClymont2025arXiv}; \textsc{FIRE-2}, \citealt{Marszewski2024ApJ}; \textsc{Serra}, \citealt{Pallottini2025AAP}).
\textit{Right}: the low-mass end of the MZR. Our observed MZR (stars) are compared with a semi-analytical model including explicit Pop III and Pop II metal enrichment \citep[\textsc{A-sloth};][]{Liu2025arXiv}. All star forming galaxies at $z\sim6$ in the simulation are plotted (black dots), and those under the first major burst of star formation are marked with blue circles.
The holizontal dashed line corresponds to the maximum metallicity that can be achieved with pure-Pop III enrichments, predicted by theoretical simulations \citep[e.g.,][]{Jaacks2018MNRAS}.
}
\label{fig:MZR}
\end{figure*}

The GLIMPSE-DDT NIRSpec observation reveals relatively strong oxygen emission lines even from the faint-end galaxies at $z\sim6$.
The $R3$ line ratios of the sample galaxies are all $R3>1.5$, which corresponds to $>1\ \%$ solar metallicity, down to the stellar mass of $10^{5.6}\ M_\odot$.
Figure~\ref{fig:MZR} left presents the mass-metallicity relation (MZR) of our sample galaxies compared with previous measurements at the similar redshift in literature \citep{Nakajima2023ApJS,Curti2024AAP,Chemerynska2024ApJ,Topping2025ApJ}.
For a fair comparison, we recompute the oxygen abundance from $R3$ using the same calibration \citep{Nakajima2022ApJS} as used in this work when the literature adopted different calibration.
The figure demonstrates that the low-mass end slope of the MZR is extended down to $\sim10^6\ M_\odot$ mass regime.

Having a good constraint on the low-mass end of the MZR from our results, we can robustly parameterize the MZR at a wide range of the stellar mass.
We adopt the analytical form of the MZR as suggested by \citet{Curti2020MNRAS}:
\begin{equation}\label{eqn:MZR}
    12+\log({\rm O/H}) = Z_0 - \frac{\gamma}{\beta}\ \log  [ 1+ (M_\star/M_0)^{-\beta} ],
\end{equation}
where $M_0$ is the characteristic turnover stellar mass above which the metallicity asymptotically reaches the saturated metallicity $Z_0$, $\gamma$ is the slope of the MZR at $M_\star<M_0$, and $\beta$ quantifies the ``width'' of the transition from the power-law MZR at $M_\star<M_0$ to the asymptotic MZR at $M_\star>M_0$.
We use the python implementation of the MCMC \citep[\texttt{emcee};][]{emcee} to derive the posterior distributions of $Z_0$, $M_0$, $\gamma$, and $\beta$ inferred from our results and metallicity measurements in literature \citep{Nakajima2023ApJS,Curti2024AAP,Chemerynska2024ApJ,Topping2025ApJ}.
A total of 131 galaxies at $z=5-8$ are used in the fit.
We obtain $Z_0=8.07^{+0.04}_{-0.03}$, $M_0=8.85^{+0.18}_{-0.13}$, $\gamma=0.35^{+0.02}_{-0.03}$, and $\beta=1.76^{+65.6}_{-0.0}$. The median MZR posterior is shown by red solid curve in Figure~\ref{fig:MZR}, with the red shaded band denotes 16th- to 84th-percentile range.

The low-mass end slope $\gamma$ is well constrained as $\gamma=0.35^{+0.02}_{-0.03}$.
In the simple framework of ``equilibrium" models \citep[e.g.,][]{Lilly2013ApJ}, this MZR slope agrees quite well with the prediction by energy-driven wind feedback \citep[$\gamma\sim0.33$; see also][for a similar discussion]{Guo2016ApJ}.
The faint-end slope of the MZR is also comparable to that in the local universe \citep[$\gamma=0.28$;][the black curve in Figure~\ref{fig:MZR} left]{Curti2020MNRAS}.
It should thus suggest a weak (redshift) evolution of the low-mass end slope of the MZR and that energy-driven outflows play a key role even in the early metal enrichment process.


The observed MZR is compared with predictions of cosmological simulations in Figure~\ref{fig:MZR} left (\textsc{FirstLight}, \citealt{Langan2020MNRAS}; \textsc{Thesan-Zoom}, \citealt{McClymont2025arXiv}; \textsc{FIRE-2}, \citealt{Marszewski2024ApJ}; \textsc{Serra}, \citealt{Pallottini2025AAP}).
The observations and predictions agree well overall.
The observed MZR has slightly shallower slope and higher normalization than cosmological simulations, but the difference is small and cannot be distinguished conclusively considering the measurement errors and possible systematic uncertainties.
Comparing the scatter around the median MZR is also of interest since different simulations predict different scatters.
The observed MZR scatter ($\sigma_Z=0.23$ dex) is larger than predictions by \textsc{FirstLight}, \textsc{Thesan-zoom}, comparable to \textsc{Serra} (with feedback modification as discussed by \citealt{Pallottini2025AAP}), and smaller than \textsc{FIRE-2}.
This could be suggestive for the gas mass fractions \citep{Langan2020MNRAS}, feedback efficiency \citep{Pallottini2025AAP}, or star-formation burstiness \citep{Marszewski2024ApJ} in cosmological simulations, though, the observed MZR scatter should be highly affected by the selection bias and systematic scatter in the $R3$-to-metallicity calibration.

\subsubsection{Two distinct pathways of the early metal enrichment?}\label{subsubsec:overshoot}
Interestingly, the lowest-mass galaxy in our sample (GLIMPSE-47757; $M_\star=10^{5.62}\ M_\odot$) shows relatively strong [O{\sc iii}]5007 line and has $R3=2.55\pm0.16$, which results in an oxygen abundance of $12+\log({\rm O/H})=7.15^{+0.03}_{-0.06}$.
Figure \ref{fig:MZR} right compares the observed MZR in this work and a semi-analytical simulation including explicit Pop III metal enrichment and subsequent Population II (Pop II) star formation and enrichment \citep[\textsc{A-sloth};][]{Liu2025arXiv}.
In the simulation, two distinct pathways of the earliest metal enrichment are predicted, which is distinguishable at $M_\star \lesssim10^6\ M_\odot$.
Galaxies under the first major burst of star formation in the simulation (i.e., more than half of stars are formed within the last 10 Myr; blue points in Figure~\ref{fig:MZR} right) typically experience either a very rapid metal enrichment up to $\sim1-10\ \%$ solar at $M_\star\sim10^{4-5}M_\odot$ (``overshoot") or a delayed enrichment where the halo is kept relatively pristine and get metal enriched with the initial starburst forming $M_\star>10^{5}M_\odot$ stars (``undershoot").

Simulations have shown that the transition from Pop III to Pop II formation phase occurs almost immediately after the galaxy experiences the first Pop III supernovae (SNe), when the halo mass (and the stellar mass) is yet small \citep[e.g.,][]{Jaacks2019MNRAS}. The early transition to the Pop II phase enables enhanced levels of early metal enrichment due to the subsequent Pop II SNe, which should be observed as the ``overshoot" population.
Under the ``overshoot" process, the metallicity goes well above the maximum metallicity that can be achieved with pure-Pop III enrichment \citep[$\sim1\ \%$ solar;][gray dashed line in Figure \ref{fig:MZR} right]{Jaacks2018MNRAS} even at $M_\star<10^5\ M_\odot$  because of the rapid Pop II SN activity following-up on the initial metal production.
This process seems to be the predominant process of the earliest chemical enrichment in the main branch, however, under some conditions, the metal enrichment is suppressed until the halo mass grows.
This could be due to strong UV feedback suppressing the star formation \citep[e.g.,][]{Jeon2014MNRAS,Sugimura2024ApJ}, inefficient metal mixing \citep[e.g.][]{Liu2020MNRAS}, or very low metal ejection from massive Pop III SNe \citep[see,][for a review]{Karlsson2013RvMP}.
When a galaxy is kept nearly pristine until the halo grows substantially, 
the galaxy experiences smoother Pop III to Pop II transition and undergoes major bursts of Pop II star formation at a high stellar mass regime. Such galaxies will be observed as the ``undershoot'' population.
These two different enrichment processes eventually merged above a high enough halo mass (or stellar mass), where the tight mass-metallicity relation starts to appear.

The relatively high metallicity of GLIMPSE-47757 ($\sim3\ \%$ solar metallicity at $M_\star=10^{5.62}\ M_\odot$) is consistent with the ``overshoot"  population in the simulation, and could indicate that the galaxy has experienced the rapid metal enrichment process.
On the other hand, recent reports of extremely low-metallicity galaxies with stellar masses of $\sim10^{5-6}\ M_\odot$ \citep[e.g.,][]{Morishita2025arXiv,Cai2025arXiv} is consistent with the other enrichment process (``undershoot") and thus it should be witnessing the delayed chemical enrichment process.
At this redshift of $z\sim6$, the two metal enrichment process seems to merge at $M_\star\gtrsim10^6\ M_\odot$, and our sample is not large enough to conclusively determine which of the two processes seem predominant observationally.
It is required to enlarge the sample of ultra-low-mass galaxies at high-$z$ ($<10^6\ M_\odot$) with metallicity measurements to establish a solid picture of the first metal enrichment process.

\subsection{Dust production in the early universe}\label{subsec:dust}
The Balmer decrements of our sample galaxies are generally consistent with the Case B recombination expectation, and most of the sample galaxies appear to be dust-free.
The only exception is GLIMPSE-55241, where we found a significant deviation from the Case B prediction in the H$\alpha$/H$\beta$ ratio ($3.01\pm0.05$; 3-sigma deviation from 2.86), which corresponds to a dust attenuation of $A_V=0.18\pm0.06$ mag.
The H$\gamma$/H$\beta$ ratio also suggests similar amount of dust attenuation ($A_V\sim0.2$ mag), and thus we infer the presence of non-negligible dust in this faint, low-mass galaxy at $z_{\rm spec}=5.5$.

\begin{figure}[t]
\centering
\includegraphics[width=0.95\linewidth]{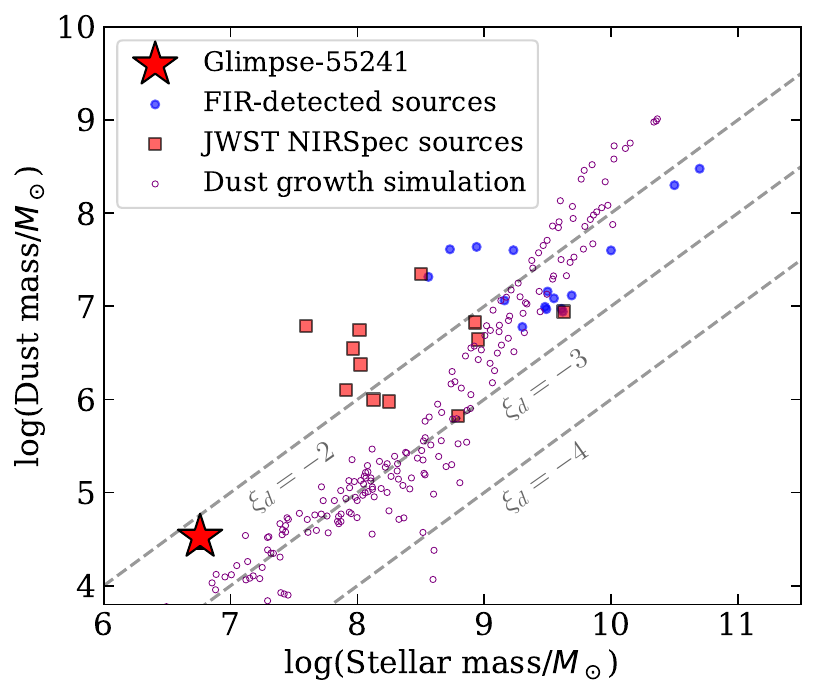}
\caption{The dust-mass vs stellar-mass plot of dust-obscured galaxies at $z>5$. GLIMPSE-55241 (red star) is compared with other FIR-detected galaxies and JWST/NIRSpec sources observed in other gravitational lensing fields. Predictions by cosmological zoom-in simulations are also plotted.
}
\label{fig:dust_mass}
\end{figure}

Following \citet{Ferrara2025AAP}, we estimate the dust mass in GLIMPSE-55241 based on $A_V$ and the size $r_e$, where
the dust mass can be estimated as $M_d=1.38\times10^5 (r_e/100\ {\rm pc})^2 A_V M_\odot$ assuming the spherical geometry.
The size of GLIMPSE-55241 is measured in the NIRCam F115W image with {\texttt Galfit} \citep{Peng2010AJ}.
Note that \citet{Casey2024ApJ} demonstrated that the dust mass estimations from $A_V$ and size give consistent results with dust masses derived from the FIR luminosity.

We obtain $M_d=10^{4.52\pm0.12}\ M_\odot$, which corresponds to a dust-to-stellar mass ratio of $\xi_d = \log(M_d/M\star)=-2.24\pm0.13$.
Note that the dust-to-stellar mass ratio is free from lens magnification uncertainty as long as stars and dust are distributed similarly.
Figure \ref{fig:dust_mass} presents the position of GLIMPSE-55241 in the $M_d$-$M_\star$ plane, comparing with other $z>5$ galaxies with dust attenuation.
We compare with FIR-detected massive galaxies \citep{Cooray2014ApJ,Watson2015Natur,Laporte2017ApJ,Hashimoto2019PASJ,Ferrara2022MNRAS,Sugahara2025ApJ,Sun2025arXiv} and JWST/NIRSpec sources with dust attenuation measured with the Balmer decrement.
For JWST/NIRSpec sources, we use NIRSpec/Prism observations in gravitational lensing fields taken by the CAnadian NIRISS Unbiased Cluster Survey (CANUCS) program \citep{Willott2022PASP,Sarrouh2025arXiv} and measure $M_d$ in the same manner.
The observed $\xi_d$ in GLIMPSE-55241 is comparable to other dust-obscured galaxies at $z>5$, while its stellar mass is remarkably lower than others.

The figure also compares the observation with a recent simulations from \citet{Narayanan2025arXiv} who model dust grain growth and processing at high-$z$.
GLIMPSE-55241 is consistent with the very first phase of dust production in those simulations, where dust production via SNe dominates over dust growth in the ISM.
However, we should note that this occurs in the simulation at $z>12$ while the GLIMPSE-55241 is at $z=5.5$.
It remains to be seen whether the $M_d$-$M_\star$ relation during the early dust production phase is redshift-dependent, though, at least in the simulation box, the $M_d$ and $M_\star$ observed in GLIMPSE-55241 are consistent with the early phase of dust production by SNe with negligible dust destruction and ISM growth.
Indeed, the observed $\xi_d$ in GLIMPSE-55241 is within the range of plausible dust yields of per SN \citep[$0.1$-$1\ M_\odot$; see, e.g.,][for a review]{Schneider2024AAPR}.
As a result, GLIMPSE-55241 could provide a snapshot into the SNe-driven dust enrichment scenarios possibly at play at very early cosmic times; though, contributions to the total dust reservoir by grain growth in the ambient ISM and other sources are not fully ruled out. Future simulations that explore grain growth and dust production from a variety of sources at the mass and redshift range of GLIMPSE-55241 are needed to unveil the origin of dust in this target.

\subsection{Ionizing photon production and escaping}\label{subsec:xi_ion}
The sample galaxies in general show very high ionizing photon production efficiencies with a large scatter.
The median $\xi_{\rm ion, obs}$ value among the entire sample is $\log(\xi_{\rm ion, obs}/{\rm Hz^{-1}\ erg})=25.92\pm0.41$.
This is comparable to or higher than measurements at similar redshifts in literature based on brighter galaxies \citep[e.g.,][]{Atek2024Natur,Harshan2024MNRAS}.
Given the median $M_{\rm UV}$ of our sample ($-16.2$ mag), this high $\xi_{\rm ion, obs}$ value is consistent with an extrapolation of the $\xi_{\rm ion, obs}$-$M_{\rm UV}$ relation obtained by \citet{Llerena2025AAP}.
The large scatter of $\xi_{\rm ion, obs}$ could be due to the systematic uncertainty in the slitloss correction (i.e., scaling factors to match NIRSpec to NIRCam). The nominal S/Ns of NIRSpec and NIRCam data estimate the scaling factors typically have $\sim0.4$ dex uncertainties, thus the scatter of $\xi_{\rm ion, obs}$ values could be mainly driven by the systematic uncertainty.

\begin{figure}[t]
\centering
\includegraphics[width=0.95\linewidth]{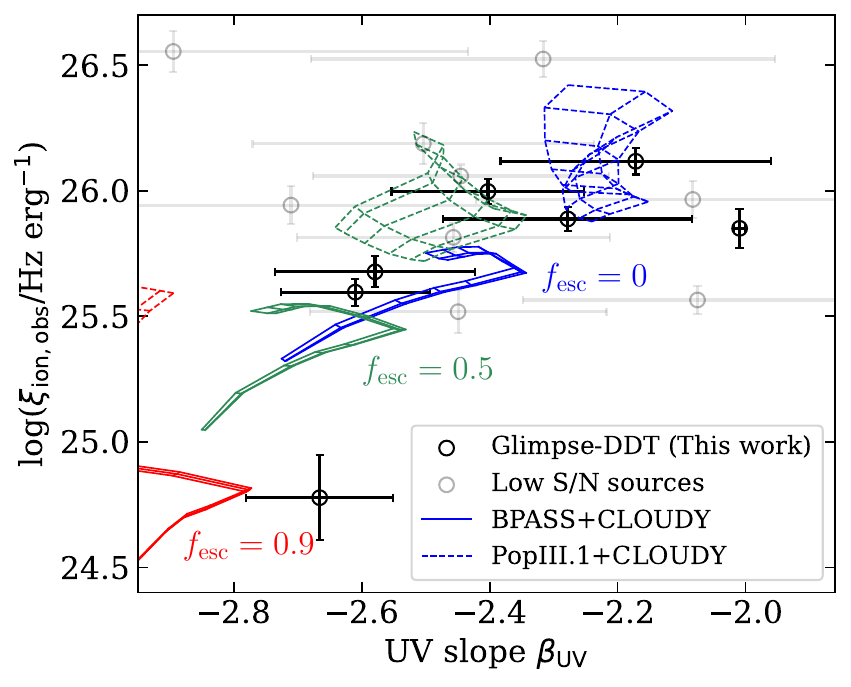}
\caption{Observed (dust-corrected but not $f_{\rm esc}$-corrected) ionizing photon production efficiency $\xi_{\rm ion, obs}$ and the rest UV slope $\beta_{\rm UV}$ of selected galaxies in this work.
High-S/N sources in rest UV NIRCam photometry are marked with thick symbols.
The photoionization model grids with \texttt{Cloudy} are also shown, varying the stellar age, gas electron density, and escape fraction. 
}
\label{fig:xi_ion_beta}
\end{figure}

Figure \ref{fig:xi_ion_beta} shows the distribution of the sample galaxies in the $\xi_{\rm iob, obs}$-$\beta_{\rm UV}$ plane.
Since we only have three or four photometric data points from NIRCam broad-band filters that we can use to measure the rest UV slope $\beta_{\rm UV}$ and its measurement is severely affected by photometric errors, we highlight sources detected at high S/Ns ($>10$) in all rest UV filters.
Although the sample size is small, higher $\xi_{\rm ion}$ sources might show less blue colors in rest UV, which could be interpreted as the significant contribution by nebular emissions (both continuum and lines). This trend is seemingly opposite to the general trend often reported in literature \citep[e.g.,][]{Bouwens2016ApJ,Matthee2017MNRAS,Saldana-Lopez2023MNRAS}. This is probably because our sample is so confined to faint-end star-forming galaxies at $z\sim6$ that there is no significant variation in the stellar population among the sample and we are mostly observing the scatter due to nebular conditions including $f_{\rm esc}$, while in literature the sample is build at a wide magnitude/redshift range and the overall trend is predominantly determined by the variation in the stellar population \citep[see also;][for the diversity of galaxy population in general at this redshift]{Jecmen2026ApJ}.

The figure also shows photoionization model grids in the $\xi_{\rm ion}$-$\beta_{\rm UV}$ diagram.
We use \texttt{Cloudy} \citep[v23.01;][]{Chatzikos2023RMxAA} to model the nebular emissions for a given stellar population.
We use \texttt{BPASS} \citep{Eldridge2017PASA} SSP as the fiducial input spectrum and vary the stellar age, gas electron density, and escape fraction to compute the grid (colored solid grids in the figure). We fix the gas metallicity to $1\ \%$ solar. 
As an extreme alternative, we also utilize PopIII SSPs with a top-heavy IMF from \texttt{Yggdrasil} \citep{Zackrisson2011ApJ} to compute the photoionization model grids (colored dashed grids).
The model grids demonstrate that standard SSP models can achieve up to $\log(\xi_{\rm iob, obs}/{\rm Hz^{-1}\ erg})\sim25.8$, and a large fraction of sample galaxies (10/16) exceeds this limit.
Instead, when PopIII SSPs with a top-heavy IMF are used as the irradiation source, it can achieve $\log(\xi_{\rm iob, obs}/{\rm Hz^{-1}\ erg})>26$ \citep[see also;][for possible non-PopIII explanation of the high $\xi_{\rm ion}$ values with Very Massive Stars]{Schaerer2025AAP}.
This indicates that a non-negligible fraction of faint-end galaxies near the end of the Epoch of Reionization can have extreme stellar populations, such as evolving (top-heavy) IMFs \citep[e.g.,][]{Mauerhofer2025AAP}, which predominantly contribute to the production of hydrogen-ionizing photons in these galaxies.
The high $\xi_{\rm ion}$ values may also be due to hidden AGNs, although no signs of broad-line components are identified.

Intriguingly, our sample includes a blue and low-$\xi_{\rm ion, obs}$ galaxy at $z_{\rm spec}=6.1$ (GLIMPSE-54637).
GLIMPSE-54637 is consistent with the locus of \texttt{BPASS+Cloudy} models in this diagram with an escape fraction of $f_{\rm esc}\sim0.9$.
Slightly evolved stellar population ($>10$ Myr) may replicate similar UV color and $\xi_{\rm ion, obs}$, but this is unlikely for GLIMPSE-54637 considering the detection of [O{\sc iii}]4959,5007 lines which requires hard ionizing photons (c.f., \citealt{Faisst2024ApJ} show that standard SSPs aged $>5$ Myrs cannot produce [O{\sc iii}] lines). 
The galaxy thus most likely produces ionizing photons intrinsically as efficiently as other UV faint galaxies ($\log(\xi_{\rm ion}/{\rm Hz^{-1}\ erg})\sim25.7$) but most of the ionizing photons should escape to the IGM and contribute to reionize the universe.

It should be noted that our sample seems to be biased towards high $\xi_{\rm ion, obs}$ galaxies.
Our sample selection requires UV faintness and the detection of H$\alpha$ or H$\beta$ with NIRSpec, which preferentially select high $\xi_{\rm ion, obs}$ sources.
To obtain the true $\xi_{\rm ion, obs}$ distribution among UV faint galaxies and evaluate their role in the cosmic reionization, a more complete sample of UV faint galaxies is needed (e.g., \citealt{Endsley2024MNRAS,Munoz2024MNRAS,Jecmen2026ApJ}, Chisholm et al. in prep.).

\section{Discussion: Roles of Supernovae}\label{sec:discussion}

\begin{figure*}[t]
\centering
\includegraphics[width=0.8\textwidth]{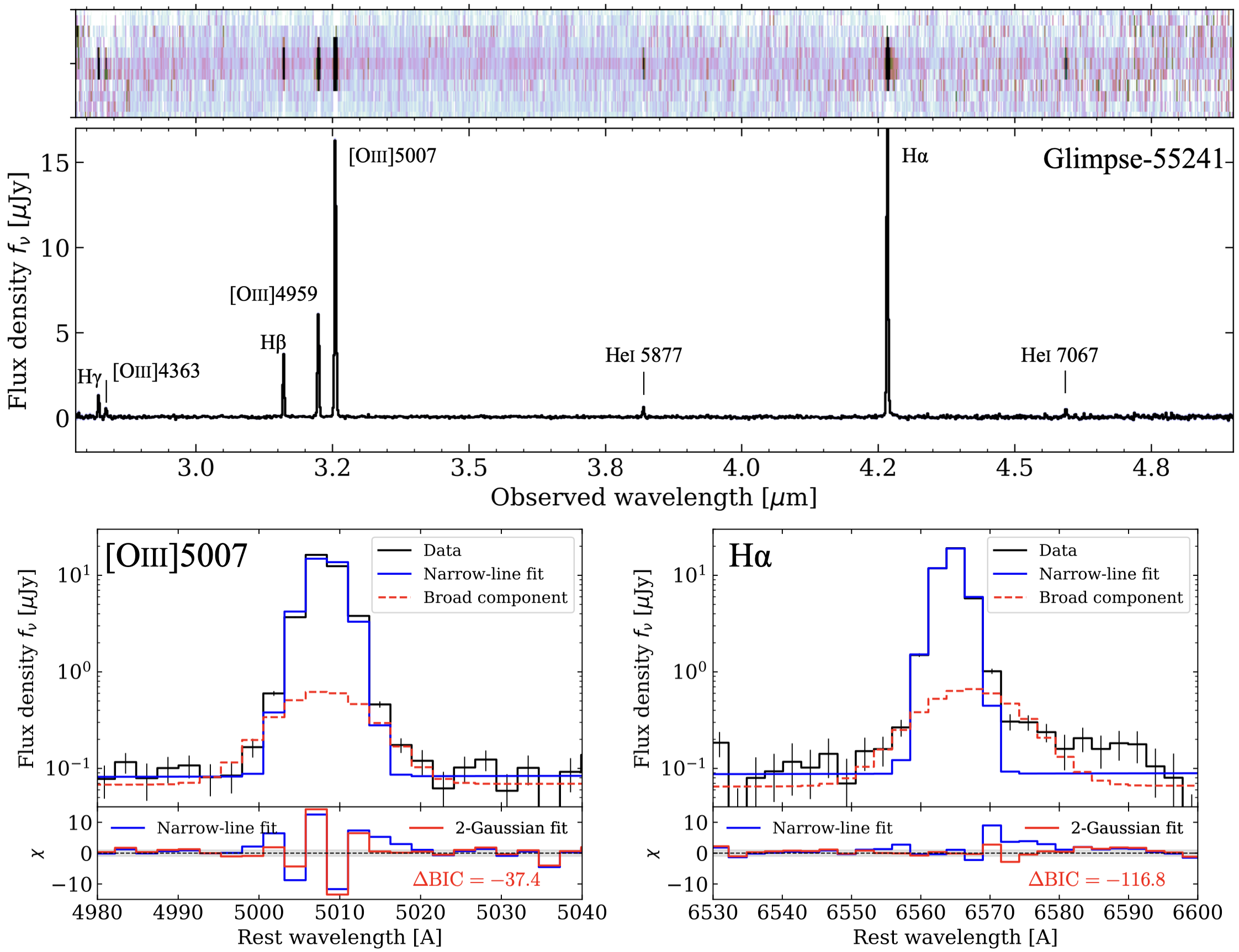}
\caption{Outflow galaxy candidate GLIMPSE-55241 among the sample. {\it Top}: The full spectrum of GLIMPSE-55241. Multiple emission lines are detected at high S/Ns.
{\it Bottom}: [O{\sc iii}]5007 and H$\alpha$ line profiles of GLIMPSE-55241. The blue lines denote the best-fit single Gaussian profiles, and the red dashed lines show the additional broad line component needed to replicate the observed line profile.
In the bottom sub-panels, the noise-normalized residuals are shown for narrow-line-only fits and two Gaussian fits.
For both [O{\sc iii}]5007 and H$\alpha$ lines, the two Gaussian models are preferred based on the  Bayesian Information Criterion.
}
\label{fig:outflow}
\end{figure*}

As discussed in Sec.~\ref{subsec:dust}, we detect dust attenuation robustly in GLIMPSE-55241, and the dust production of this galaxy is expected to be mainly driven via the efficient dust production by SNe.
Along with the dust detection, most interestingly, we identify the ionized gas outflow from the same galaxy.
Therefore GLIMPSE-55241 galaxy provides an unprecedented opportunity to probe the role of SNe in various aspects of early galaxy evolution, including dust production, metal enrichment, and stellar feedback.
In this section, we explore the detailed properties of GLIMPSE-55241 and connect its dust content, metals, and outflow through SNe.

\subsection{Outflow properties in GLIMPSE-55241}\label{subsec:outflow}
The gas outflow in GLIMPSE-55241 is identified with the broad line component both in the H$\alpha$ and [O{\sc iii}]5007 line profiles (Figure \ref{fig:outflow}).
The observed line profiles of H$\alpha$ and [O{\sc iii}]5007 cannot be fitted well with a single Gaussian leaving significant residuals in both lines, and two-Gaussian models are preferred based on the Bayesian Information Criteria.
We can infer the ionized-gas outflow properties based on the two-Gaussian profile fittings on the GLIMPSE-55241 spectrum.
Following \citet{Carniani2024AAP}, we measured the outflow velocity as
\begin{equation}
    v_{\rm out} = |v_{\rm broad} - v_{\rm narrow}| + 2\sigma_{\rm broad,deconv},
\end{equation}
where the first term is the velocity offset between the narrow and broad component, and the second component is the (line-spread-function-corrected) line width of the broad component.
We also estimated the outflow gas mass from the broad line luminosities
\begin{equation}\label{eqn:outflow_Ha}
    M_{\rm out}^{\textrm{H}\alpha} = 0.8\times10^5 \left(\frac{ L_{ \textrm{H} \alpha}}{10^{40} {\rm erg\ s^{-1}}} \right) \left(\frac{n_{\rm out}}{100\ {\rm cm^{-3}}} \right)^{-1}\ M_\odot,
\end{equation}
and 
\begin{dmath}\label{eqn:outflow_O3b}
    M_{\rm out}^{\textrm{[OIII]}} = 0.8\times10^8 \\\left(\frac{ L_{ \textrm{[OIII]}}}{10^{44} {\rm erg\ s^{-1}}} \right) \left(\frac{Z_{\rm out}}{Z_\odot} \right)^{-1} \left(\frac{n_{\rm out}}{500\ {\rm cm^{-3}}} \right)^{-1}\ M_\odot,
\end{dmath}
where $Z_{\rm out}$ and $n_{\rm out}$ are the metallicity and the electron density of the outflow gas, respectively.
We assume $n_{\rm out}=380\ {\rm cm^{-3}}$ following the assumption in literature \citep[e.g.,][]{Carniani2024AAP,Ivey2025arXiv} and $Z_{\rm out}=0.07 Z_\odot$ from the metallicity measurement in Sec.~\ref{subsec:prop}.

\begin{deluxetable}{cccc}
    \label{tab:outflow_params}
    \tablecaption{Outflow properties in a $M_\star=10^{6.6}\ M_\odot$ galaxy at $z=5.5$, GLIMPSE-55241.
  	}
    \tablewidth{\textwidth}
    \tablehead{
    \colhead{Line} & \colhead{$v_{\rm out}$ [km s$^{-1}$]} & \colhead{$\log(M_{\rm out}/M_\odot)$} & \colhead{$\eta$}
    \\
    \colhead{(1)} & \colhead{(2)} & \colhead{(3)} & \colhead{(4)}
    }
    \startdata
    H$\alpha$ & $690\pm61$ & $4.62\pm0.04$ &$0.7\pm0.1$ \\
    \ [O{\sc iii}]5007 & $611^{+66}_{-58}$ & $5.58\pm0.06$ & $6.2\pm0.6$ \\
    Fiducial & $650\pm40$ & $5.1\pm0.5$ & $2.1^{+4.1}_{-1.4}$ \\
    \enddata
    \tablenotetext{}{(1) Line used to compute the outflow properties. (2) Outflow velocity. (3) Outflow gas mass. (4) Mass loading factor.
    }
\end{deluxetable}

Table \ref{tab:outflow_params} present the outflow properties.
The outflow velocities measured with H$\alpha$ and [O{\sc iii}]5007 lines agree with each other, and the outflow velocity is faster than the median value but within the scatter reported in literature at this redshift \citep[e.g.,][]{Ubler2023AAP,Carniani2024AAP}.
On the other hand, the outflow gas masses estimated from H$\alpha$ and [O{\sc iii}]5007 lines differ by a factor of $\sim1$ dex, which could be due to systematic uncertainties in the two calibrations of Eqn.~(\ref{eqn:outflow_Ha}) and (\ref{eqn:outflow_O3b}).
For example, the metallicity of the outflow gas might be more metal enriched than the system ISM \citep[e.g.,][]{Chisholm2018MNRAS}, and assuming higher metallicity for the outflow gas would decrease the $M_{\rm out}^{\textrm{[OIII]}}$ and relieve the discrepancy.
We therefore take the mean of $\log(M_{\rm out}^{\textrm{H}\alpha})$ and $\log(M_{\rm out}^{\textrm{[OIII]}})$ as the fiducial value of the outflow mass.
Overall, the outflow found in GLIMPSE-55241 has a velocity of $v_{\rm out}=650\pm40\ {\rm km\ s^{-1}}$ and the gas mass is $M_{\rm out}=10^{5.1\pm0.5} M_\odot$.
The fiducial outflow mass and velocity correspond to a mass loading factor of $\eta=2.1^{+4.1}_{-1.4}$, which is comparable to what is typically found in literature at the similar redshift \citep[e.g.,][]{Carniani2015AAP, Ivey2025arXiv}.




\subsection{Connecting the dots -- dust, outflow, and metal enrichment regulated by SNe}\label{subsec:55241}

\noindent
$\bullet$ \underline{Energy balance from the dust and outflow via SNe}
\vspace{4pt}

Galactic outflows particularly in low-mass galaxies are expected to be primarily powered by SNe \citep[e.g.,][]{Silk2012RAA,vonGlasow2013MNRAS}.
To connect the dust production and observed outflow in GLIMPSE-55241, we examine the energy budget expected for SNe ever occurred in this galaxy.
We first estimate the total number of SNe from the stellar mass estimation.
Assuming the Chabrier IMF, the total number of SNe ever occurred in GLIMPSE-55241 is estimated as $N_{\rm SN}\sim10^{4.8}$. 
Given a dust yield of $m_d=0.5\ M_\odot$ per one SN event \citep[][]{Schneider2024AAPR}, the total dust mass predicted from the number of SNe is $\sim10^{4.5}\ M_\odot$, which agrees well with the dust mass estimation from the Balmer decrement and size measurement presented in Sec.~\ref{subsec:dust}.
On the other hand, considering that the energy released by SNe ($\sim10^{51}\ {\rm erg}$ per one SN) is typically converted into kinetic energy of galactic outflows with 1-10\ \% efficiency \citep[e.g.,][]{vonGlasow2013MNRAS,Veilleux2005ARAA}, the outflow kinetic energy driven by the SNe is predicted as $E_{\rm kin, out}\sim10^{53.8-54.8}\ {\rm erg}$.
This energy is fully consistent with the observed kinetic energy of the gas outflow estimated as $M_{\rm out}v_{\rm out}^2\sim10^{54.0}$ erg.
The energy budget agreement between dust and outflow in GLIMPSE-55241 indicates that they can be consistently explained by SNe.

\noindent
$\bullet$ \underline{Dust and oxygen enrichment by SNe}
\vspace{4pt}

In addition to dust, (core-collapse) SNe are expected to be the major origin of oxygen enrichment in the universe \citep[see;][for a review]{Kobayashi2020ApJ}.
To bridge the origin of dust and oxygen detected in the spectrum of GLIMPSE-55241, we compare the dust mass and oxygen mass in the system and examine the mass budget.

We first compute the total amount of oxygen produced by the number of SNe predicted from the stellar mass measurement ($N_{\rm SN}\sim10^{4.8}$).
Given the Chabrier03 IMF \citep{Chabrier2003PASP} and oxygen ejecta mass from CCSNe \citep{Portinari1998AAP}, a total of $\sim4\ M_\odot$ oxygen atoms are produced and ejected per one SN event. The GLIMPSE-55241 galaxy is thus expected to have produced $\sim10^{5.4}\ M_\odot$ of oxygen atoms in total so far.

We next estimate the oxygen mass ($M_{\rm Oxy}$) in the galaxy ISM and outflow from the line luminosity with a similar approach as Eqn (\ref{eqn:outflow_O3b}).
Given the different ionized gas conditions in the ISM and outflow gas, we use different conversion factors from line luminosity to oxygen mass for ISM and outflow component (conversion factors are derived in Appendix \ref{apx:O3b_convert}):
\begin{align}
    M_{\rm Oxy}^{\rm ISM} =&\ 1.3\times10^5 \left(\frac{L_{\rm [OIII]}^{\rm narrow}}{10^{44}\ {\rm erg\ s^{-1}}}\right) \left(\frac{n_e}{1000\ {\rm cm^{-3}}} \right)^{-1} \ M_\odot,\label{eqn:Oxymass_ism} \\
    M_{\rm Oxy}^{\rm out} =&\ 9.6\times10^5 \left(\frac{L_{\rm [OIII]}^{\rm broad}}{10^{44}\ {\rm erg\ s^{-1}}}\right) \left(\frac{n_e}{500\ {\rm cm^{-3}}} \right)^{-1} \ M_\odot. \label{eqn:Oxymass_out}
\end{align}
Note that this conversion does not depend on metallicity and is most affected by electron temperature and density (see Appendix \ref{apx:O3b_convert}).
We obtain the oxygen mass in ISM as $M_{\rm Oxy}^{\rm ISM}=10^{2.6}\ M_\odot$ and that in outflow as $M_{\rm Oxy}^{\rm out}=10^{2.5}\ M_\odot$.

These oxygen mass budget estimations suggest that SNe produce enough amount of oxygen and considerable amount of them seems to be expelled presumably due to galactic outflows.
Observations of low-mass dwarfs in the local universe have shown that they can retain only a few percent of metals ever produced and most of them seems to be lost due to stellar feedback \citep[e.g.,][]{Kirby2011ApJ,McQuinn2015ApJ}.
Our observations are consistent that, in the order of magnitude, only $\sim0.1-1\ \%$ of oxygen should be retained in the galaxy ISM, and significant amount of oxygen is being removed through the outflow.

We note that the oxygen mass estimation from the [O{\sc iii}] line luminosity can have systematic uncertainties, but it unlikely affects the conclusion here.
The oxygen mass in the ISM phase ($M^{\rm ISM}_{\rm Oxy}$) can be underestimated by a factor of $\sim2-3$ at most, while the oxygen mass in the outflow phase ($M^{\rm out}_{\rm Oxy}$) can be more likely to be underestimated (see Appendix \ref{apx:O3b_convert} for the detail).
It is thus the fraction of oxygen retained in the ISM is expected to be no larger than a few percent even when possible systematic uncertainties are accounted.

\noindent
\underline{$\bullet$ Tying altogether: early galaxy assembly with SNe}
\vspace{4pt}

Owing to the unique ensemble of the successful detection of the dust attenuation and the outflow in an ultra-faint galaxy at $z>5$,
GLIMPSE-55241 gives the first insight into the key roles of SNe in early galaxy formation.
It observationally suggest that SNe can be the primary source of dust and metal enrichment in a low-mass galaxy at $z=5.5$.
The SNe can also power the galactic outflow, which could remove considerable amount of metals from the ISM and release them to CGM or IGM.
Such feedback-driven metal enrichment in the galaxy ISM is consistent with our result of the low-mass end slope of the MZR (Sec.~\ref{subsec:MZR}).
SNe thus seem to play a crucial role in shaping the early galaxy evolution on various aspects including dust production, metal enrichment, stellar feedback, and baryon cycle between ISM and CGM/IGM, as predicted from simulations and lower-$z$ observations \citep[e.g.,][]{Silk2012RAA, Tumlinson2017ARAA,Smith2018MNRAS,Narayanan2025arXiv}.

We should caution that the discussion here is based on very rough estimations, 
where major systematic uncertainties may remain. 
In particular, the dust mass obtained in Sec.~\ref{subsec:dust} could be overestimated, since the formulation by \citet{Ferrara2025AAP} implicitly assumes that the spatial distribution of dust and stars are similar and the size $r_e$ is measured from the NIRCam image. If the dust distribution is more compact, the dust mass measurement would decrease proportionally to the square of the size.
If the dust mass is overestimated, then the dust production predicted from the number of CCSNe would be larger than what is observed in GLIMPSE-55241 --- in this case, dust removal from the ISM via galactic outflows could account for the missing dust.
Nevertheless, the order-level consistency between dust, outflow, and oxygen abundance connected via SNe in GLIMPSE-55241 offers an unprecedented opportunity to observationally investigate 
the roles of SNe on the early galaxy formation.


\section{Conclusion}\label{sec:conclusion}
Spectroscopic characterization of faint-end galaxies at $z>5$ is crucial to understand the earliest galaxy formation.
In particular, a statistic overview of their spectroscopic properties has been missing due to the lack of a large sample of highly magnified, very faint galaxies with deep NIRSpec observations.
In the paper, we leverage the deep NIRCam and NIRSpec observations in a gravitational lensing field of Abell S1063, taken as part of the GLIMPSE \citep{Atek2025arXiv} survey and the GLIMPSE-D program (PID:9223, PIs: S.~Fujimoto \& R.~Naidu), and obtain a sample of 16 galaxies fainter than $M_{\rm UV}=-17$ mag at $5.5\lesssim z \lesssim6.5$.
The lens magnification enable us to explore the physical properties down to $M_{\rm UV}=-14.6$ mag (or $M_\star=10^{5.6}\ M_\odot$) galaxies with detecting the key rest optical emission lines such as H$\alpha$, H$\beta$, and [O{\sc iii}]4959,5007.
Our main findings are:

\begin{enumerate}
    \item We identify relatively strong [O{\sc iii}]4959,5007 lines from all sample galaxies despite of their faintness. Based on their [O{\sc iii}]/H$\beta$ ratios, we find that the low-mass end of the mass-metallicity relation (MZR) is extended down to $M_\star\sim10^6\ M_\odot$ (Sec.~\ref{subsec:MZR}).
    \item The low-mass end slope of the MZR is found to be $\gamma=0.35^{+0.02}_{-0.03}$, which agrees well with cosmological simulations and that of the local MZR. Under the simple ``equilibrium" model framework, the slope should suggest that the early chemical enrichment is primarily regulated by the energy-driven feedback (Sec.~\ref{subsec:MZR}, Fig.~\ref{fig:MZR} left).
    \item Even the faintest galaxy in the sample also shows a relatively high [O{\sc iii}]/H$\beta$ ratio with the stellar mass of $M_\star=10^{5.6}\ M_\odot$. Combined with recent reports of extremely metal-poor galaxy candidates at the similar stellar mass and redshift, we are perhaps witnessing two distinct pathways of the earliest chemical enrichment (overshoot vs undershoot), as predicted by
    cosmological simulations involving Pop III metal enrichment (Sec.~\ref{subsubsec:overshoot}, Fig.~\ref{fig:MZR} right).
    \item Our sample galaxies generally show quite high ionizing photon production efficiency ($\langle \xi_{\rm ion,\ obs}\rangle=10^{25.92}\ {\rm Hz^{-1}\ erg}$). The vast majority of sample galaxies exceeds the maximum $\xi_{\rm ion,\ obs}$ value that standard stellar population can achieve, suggesting that a non-negligible fraction of faint-end galaxies near the end of Epoch or Reionization should have extreme stellar populations that predominantly contribute to the ionizing-photon production in these galaxies (Sec.~\ref{subsec:xi_ion}, Fig.~\ref{fig:xi_ion_beta}).
    \item We detect the dust attenuation in one of the sample galaxies, by finding 3-sigma deviation of the Balmer decrement from the case B prediction. Considering its stellar mass and dust mass ($M_\star=10^{6.6}\ M_\odot$ and $M_d=10^{4.5}\ M_\odot$), the dust observed in this low-mass galaxy seems to mainly be produced by efficient dust production via supernovae (SNe; Sec.~\ref{subsec:dust}, Fig.~\ref{fig:dust_mass}).
    \item Moreover, the same galaxy shows ionized gas outflow identified by the broad line component in the [O{\sc iii}]5007 and H$\alpha$ lines (Sec.~\ref{subsec:outflow}, Fig.~\ref{fig:outflow}). By comparing the dust, metal, and outflow contents in this galaxy, we conclude that SNe play the crucial role in the early galaxy assembly -- the low-mass galaxy has been metal-enriched via SNe, which also produce dust and power the galactic outflow. A significant amount of metals seems to be removed by the outflow, and only a fraction of metals should be retained in the low-mass galaxy. Such SNe-driven feedback processes seem to be the key physics behind the early galaxy evolution on various aspects (Sec.~\ref{subsec:55241}).
\end{enumerate}

Our work uniquely shows that the deep NIRSpec census of faint-end galaxies with gravitational lensing effect is crucial in understanding the earliest galaxy evolution.
Not only by estimating the gas-phase metallicity of lowest-mass galaxies but by detecting the dust obscuration and/or ionized gas outflow, we can observationally infer the fundamental physics behind the galaxy evolution.
However, the sample size is not yet large enough particularly at the $M_\star<10^6\ M_\odot$ mass range to conclusively explore the first metal enrichment process, where the simulations predict different pathways of chemical enrichment.
Further NIRSpec census of highly magnified galaxies are needed.

\begin{acknowledgments}
We thank Pratika Dayal at Canadian Institute for Theoretical Astrophysics for a fruitful discussion to improve the quality of this paper.
We also thank the anonymous referee for constructive comments and suggestion to improve the quality of the paper.
This work is based on observations made with the NASA/ESA/CSA JWST. The data were obtained from the Mikulski Archive for Space Telescopes at the Space Telescope Science Institute, which is operated by the Association of Universities for Research in Astronomy, Inc., under NASA contract NAS5-03127 for JWST.
This research used the Canadian Advanced Network For Astronomy Research (CANFAR) operated in partnership by the Canadian Astronomy Data Centre and The Digital Research Alliance of Canada with support from the National Research Council of Canada, the Canadian Space Agency, CANARIE and the Canadian Foundation for Innovation.
The Dunlap Institute is funded through an endowment established by the David Dunlap family and the University of Toronto.
We acknowledge the support of the Canadian Space Agency (CSA) [25JWGO4A06]. 
YA is supported by the Dunlap Institute and JSPS KAKENHI Grant Number 23H00131.
HA acknowledge support from CNES, focused on the JWST mission, and the Programme National Cosmology and Galaxies (PNCG) of CNRS/INSU with INP and IN2P3, co-funded by CEA and CNES. IC acknowledges funding support from the Initiative Physique des Infinis (IPI), a research training program of the Idex SUPER at Sorbonne Universit\'e. HA acknowledges support by the French National Research Agency (ANR) under grant ANR-21-CE31-0838. This work has made use of the \texttt{CANDIDE} Cluster at the \textit{Institut d'Astrophysique de Paris} (IAP), made possible by grants from the PNCG and the region of Île de France through the program DIM-ACAV+, and the Cosmic Dawn Center and maintained by S. Rouberol.

\end{acknowledgments}

\section*{Data Availability}

The JWST data used in this paper were obtained from the Mikulski Archive for Space Telescopes (MAST) at the Space Telescope Science Institute.
Data can be found at \dataset[doi: 10.17909/ygrj-m815]{http://dx.doi.org/10.17909/ygrj-m815}.





%
\facilities{HST (ACS and WFC3), JWST (NIRCam and NIRSpec)}

\software{astropy \citep{2013A&A...558A..33A,2018AJ....156..123A,2022ApJ...935..167A},  
          Cloudy \citep{2013RMxAA..49..137F}, 
          Source Extractor \citep{Bertin1996AAPS},
          Pyneb \citep{Luridiana2015AAP},
          Photutils \citep{Bradley2020zndo},
          msaexp \citep{Brammer2022zndo}
          }


\appendix

\section{Emission line flux measurements}\label{apx:line_fluxes}

We list the emission line fluxes of all sample galaxies, measured in Sec.~\ref{subsec:prop}, in Table \ref{tab:sample_apx}.
We report the observed emission line fluxes in Table \ref{tab:sample_apx}, and values in the table are not corrected for gravitational lensing, dust attenuation, nor slit losses.
The correction factors of slit losses estimated by comparing with NIRCam photometry is also listed in the table.

\begin{deluxetable}{lcccccc}
    \label{tab:sample_apx}
    \tablecaption{Emission line fluxes of the UV faintest galaxies.
  	}
    \tablewidth{0pt}
    \tablehead{
    \colhead{ID} & \colhead{$z_{\rm spec}$} & \colhead{$F_{\textrm{H}\alpha}$} & \colhead{$F_{\textrm{H}\beta}$} & \colhead{$F_{\textrm{[OIII]4959}}$} & \colhead{$F_{\textrm{[OIII]5007}}$} & \colhead{Scale factor} \\
    \colhead{} & \colhead{}& \colhead{$10^{-19}$ cgs} & \colhead{$10^{-19}$ cgs} & \colhead{$10^{-19}$ cgs} & \colhead{$10^{-19}$ cgs} & \colhead{}\\
    \colhead{(1)} & \colhead{(2)}& \colhead{(3)} & \colhead{(4)} & \colhead{(5)} & \colhead{(6)} & \colhead{(7)}
    }
    \startdata
    7685 & $6.4184 \pm 0.0004$ & $2.12 \pm 0.35$ & $0.77 \pm 0.19$ & $0.84 \pm 0.18$ & $2.64 \pm 0.22$ & $2.88$ \\
8139 & $6.2233 \pm 0.0003$ & $3.94 \pm 0.37$ & $<0.8$ & $1.95 \pm 0.21$ & $5.44 \pm 0.25$ & $1.57$ \\
9081 & $6.2218 \pm 0.0002$ & $10.96 \pm 1.00$ & $3.54 \pm 0.72$ & $6.13 \pm 0.68$ & $21.00 \pm 0.80$ & $0.73$ \\
11789 & $6.5210 \pm 0.0007$ & --\tablenotemark{$\star$} & $3.72 \pm 0.69$ & $4.80 \pm 0.56$ & $9.84 \pm 0.62$ & $1.06$ \\
13186 & $6.1572 \pm 0.0002$ & $7.24 \pm 0.61$ & $2.38 \pm 0.45$ & $1.01 \pm 0.31$ & $3.65 \pm 0.39$ & $2.81$ \\
17070 & $6.0297 \pm 0.0001$ & $5.59 \pm 0.37$ & $1.63 \pm 0.32$ & $1.26 \pm 0.27$ & $3.70 \pm 0.33$ & $2.30$ \\
19034 & $6.2027 \pm 0.0002$ & $2.91 \pm 0.17$ & $1.06 \pm 0.12$ & $0.75 \pm 0.10$ & $1.79 \pm 0.12$ & $1.24$ \\
35552 & $6.5018 \pm 0.0007$ & $5.16 \pm 1.03$ & --\tablenotemark{$\star$} & $1.75 \pm 0.53$ & $6.32 \pm 0.63$ & $0.48$ \\
35957 & $6.1070 \pm 0.0001$ & $15.30 \pm 0.98$ & $5.70 \pm 0.75$ & $8.95 \pm 0.87$ & $26.56 \pm 1.08$ & $1.12$ \\
37914 & $6.1072 \pm 0.0002$ & $8.03 \pm 0.63$ & $3.54 \pm 0.57$ & $3.92 \pm 0.46$ & $10.29 \pm 0.54$ & $1.97$ \\
45084 & $6.4940 \pm 0.0003$ & $1.65 \pm 0.18$ & $0.78 \pm 0.14$ & $0.55 \pm 0.11$ & $1.81 \pm 0.13$ & $2.71$ \\
45706 & $6.4153 \pm 0.0003$ & $6.84 \pm 0.75$ & $3.10 \pm 0.49$ & $1.32 \pm 0.42$ & $5.48 \pm 0.51$ & $2.52$ \\
46431 & $6.5224 \pm 0.0006$ & $4.79 \pm 0.78$ & $1.50 \pm 0.49$ & $1.72 \pm 0.35$ & $3.34 \pm 0.40$ & $3.04$ \\
47757 & $6.0019 \pm 0.0001$ & $12.77 \pm 0.31$ & $4.75 \pm 0.27$ & $5.12 \pm 0.24$ & $12.12 \pm 0.29$ & $1.22$ \\
54637 & $6.0654 \pm 0.0006$ & $3.44 \pm 0.73$ & $<1.6$ & $1.17 \pm 0.36$ & $3.55 \pm 0.43$ & $0.60$ \\
55241 & $5.4996 \pm 0.0000$ & $107.58 \pm 0.62$ & $35.12 \pm 0.51$ & $67.17 \pm 0.57$ & $174.93 \pm 0.89$ & $0.92$ \\
    \enddata
    \tablenotetext{}{(1) Source IDs in the GLIMPSE catalog. (2) Spectroscopic redshifts. (3) H$\alpha$ line fluxes. (4) H$\beta$ line fluxes. (5) [O{\sc iii}]4959 line fluxes. (6) [O{\sc iii}]5007 line fluxes. (7) Scale factor to match to the NIRCam photometry. This correction is not applied to the fluxes in this table.
    }
    \tablenotetext{}{$\star$ The line falls in the gap of NIRSpec detectors.}
\end{deluxetable}

\section{Conversion from [OIII]5007 line luminosity to oxygen mass}\label{apx:O3b_convert}

In this section we derive the conversion factors to estimate the oxygen mass from the [O{\sc iii}]5007 line luminosity (Eqn.~\ref{eqn:Oxymass_ism} and \ref{eqn:Oxymass_out}).
We start from the line luminosity definition;
\begin{equation*}
    L_{\rm [OIII]} = \int_V f n_e n({\rm O^{2+}}) j_{\rm [OIII]}(n_e, T_e) dV,
\end{equation*}
where $f$ is the filling factor, $n_e$ is the electron density, $n({\rm O^{2+}})$ is the number density of doubly-ionized oxygen atoms, and $j_{\rm [OIII]}$ is the emissivity of the [O{\sc iii}]5007 line.
Considering that we are observing H{\sc ii} regions in a high-$z$ star-forming galaxy, we can proximate $n({\rm H})\sim n_e$ and $n({\rm O^{2+}})\sim n({\rm O})$, which lead to
\begin{equation*}
    n({\rm O^{2+}}) \sim \frac{n(O)}{n(H)} n_e = 6.04\times10^4 \times 10^{[{\rm O/H}]-[{\rm O/H}]_\odot} n_e.
\end{equation*}
This assumption is discussed later.
With this assumption, the line luminosity can be expressed as
\begin{equation}
    L_{\rm [OIII]} = 6.04\times10^4 \times 10^{[{\rm O/H}]-[{\rm O/H}]_\odot} j_{\rm [OIII]}f\langle n_e^2\rangle V.\label{eqn:B1}
\end{equation}

Let $Z_{\rm Oxy}$ be the oxygen mass fraction in the total gas mass $M_{\rm gas}$, then the oxygen mass is
\begin{align}
    M_{\rm Oxy} &= Z_{\rm Oxy} M_{\rm gas} \notag \\
    &= Z_{\rm Oxy, \odot} \times 10^{[{\rm O/H}]-[{\rm O/H}]_\odot} M_{\rm gas}. \label{eqn:B2}
\end{align}
The gas mass $M_{\rm gas}$ can be written as
\begin{equation}
    M_{\rm gas} \sim \int_Vf \bar{m} n(H) dV \sim 1.3 f m_p  \langle n_e \rangle V, \label{eqn:B3}
\end{equation}
where $\bar{m}$ is the average molecular mass of the gas and $m_p$ is the proton mass. Taking account for the $\sim10$ \% number density of He atoms w.r.t. H atoms,
\begin{equation*}
    \bar{m} \sim \frac{m_p n({\rm H}) + 4m_p n({\rm He})}{n({\rm H})+n({\rm He})} \sim \frac{1.4}{1.1}m_p \sim 1.3m_p.
\end{equation*}
Combining Eqn.~(\ref{eqn:B1}), (\ref{eqn:B2}), and (\ref{eqn:B3}), we obtain
\begin{equation}
    L_{\rm [OIII]} = 5.0\times10^{-2} \frac{M_{\rm Oxy}}{m_p}C \langle n_e \rangle j_{\rm [OIII]},
\end{equation}
where $C=\langle n_e \rangle^2/\langle n_e^2 \rangle$ is the clumping factor.

The [O{\sc iii}]5007 emissivity $j_{\rm [OIII]}$ is a function of electron density and temperature, and it can be widely different between the galaxy ISM and outflow gas.
For the ISM component, we need to assume the electron density but can measure the electron temperature from $RO3=$[O{\sc iii}]4959+5007/[O{\sc iii}]4363 line ratio in GLIMPSE-55241.
We assume $n_e=1000\ {\rm cm^{-3}}$ in line with recent works on the redshift evolution of $n_e$ \citep[e.g.,][]{Isobe2023ApJ,Abdurro'uf2024ApJ,Topping2025MNRAS}, but $n_e$ has only negligible effect on $j_{\rm [OIII]}$ as long as it is below the critical density ($n_e\sim10^6\ {\rm cm^{-3}}$).
The dust-corrected line ratio is $RO3=34.9\pm6.4$, which corresponds to $T_e=2.2^{+0.4}_{-0.3}\times10^4$ K.
Given these $n_e$ and $T_e$, the [O{\sc iii}]5007 line emissivity is $j_{\rm [OIII]}=1.32\times10^{-20}\ {\rm erg\ s^{-1}\ cm^3}$.
Assuming $C=1$, we finally obtain
\begin{equation}\label{eqn:B5}
    M_{\rm Oxy}^{\rm ISM} = 1.3\times10^5 \left(\frac{L_{\rm [OIII]}^{\rm narrow}}{10^{44}\ {\rm erg\ s^{-1}}}\right) \left(\frac{n_e}{1000\ {\rm cm^{-3}}} \right)^{-1} M_\odot.
\end{equation}
We caution that this conversion factor explicitly depends on the electron temperature. When the temperature is lower than GLIMPSE-55241, the [O{\sc iii}]5007 emissivity gets lower, and the conversion factor in Eqn.~(\ref{eqn:B5}) gets larger.
For example, when $T_e=10000$ K is assumed, the coefficient term in Eqn.~(\ref{eqn:B5}) becomes $4.8\times10^5$.

For the outflow component, it is more difficult to know the actual ionized gas conditions as the outflow is detected only in H$\alpha$ and [O{\sc iii}]5007 lines.
We therefore follow prescription by \citet{Carniani2015AAP} and assume $T_e=10000$ K and $n_e=380\ {\rm cm^{-3}}$.
With this assumption, the [O{\sc iii}]5007 emissivity is $j_{\rm [OIII]}=3.5\times10^{-21}\ {\rm erg\ s^{-1}\ cm^3}$, and we obtain
\begin{equation}\label{eqn:B6}
    M_{\rm Oxy}^{\rm out} = 9.6\times10^5 \left(\frac{L_{\rm [OIII]}^{\rm broad}}{10^{44}\ {\rm erg\ s^{-1}}}\right) \left(\frac{n_e}{500\ {\rm cm^{-3}}} \right)^{-1} M_\odot.
\end{equation}

As mentioned above, we proximate $n({\rm O}^{2+})\sim n({\rm O})$ to derive the conversion factors.
This is likely the case for H{\sc ii} regions with high ionization parameters, but does not hold either when $n({\rm O^+})$ contribution is not negligible or when we want to include the oxygen mass in H{\sc i} gas clouds.
Thus, what we can measure from Eqn.~(\ref{eqn:B5}) or (\ref{eqn:B6}) is the oxygen mass in H{\sc ii} clouds (with an ionization correction if needed), and the oxygen mass in H{\sc i} clouds needs to be taken account additionally.

For ISM in high-$z$ low-mass SFGs, the missing component can be negligible and Eqn.~(\ref{eqn:B5}) should not be a bad approximation.
JWST observations have found these galaxies typically show high ionization parameters \citep[e.g.,][]{Reddy2023ApJ}, suggesting most of oxygen atoms in H{\sc ii} region exist as the O$^{2+}$ state.
Although oxygen mass in the H{\sc i} clouds is unknown, given their compact morphology and high sSFR in general, a large fraction of hydrogen is expected to ionized in the ISM \citep[see e.g.,][for the sizes of ionized bubbles]{Morishita2023ApJ}.
\citet{Nakazato2025arXiv} showed that ionized-to-neutral gas mass ratios in the cosmological simulations can be $\sim1$ within $2R_{\rm vir}$ during the starburst phase in early galaxy formation.
It is thus the oxygen mass from Eqn.~(\ref{eqn:B5}) can be underestimated by a factor of $\sim2-3$ at most, for high-$z$ low-mass galaxies.

On the other hand, for the outflow gas, it is not clear how much oxygen mass can be underestimated with Eqn.~(\ref{eqn:B6}).
The ionization parameters in (ionized) outflow gas may be lower than ISM while non-stellar origin excitation could enhance the $n({\rm O^{2+}})/n({\rm O^{+}})$ ratio.
Moreover, neutral gas outflow component can have significant mass. E.g., \citet{Belli2024Natur} showed neutral gas mass is $\sim2$ dex higher than ionized gas mass in the outflow from a massive quiescent galaxy at $z=2.45$.
It is therefore the oxygen mass from Eqn.~(\ref{eqn:B6}) in outflow gas component could be underestimated by as much as $\sim2$ dex in some cases.

Overall, $M_{\rm Oxy}^{\rm ISM}$ by Eqn.~(\ref{eqn:B5}) is expected to be a good approximation while $M_{\rm Oxy}^{\rm out}$ by Eqn.~(\ref{eqn:B6}) may be underestimated considerably.
However, this does not affect our conclusion in Sec.~\ref{subsec:55241}.
The fraction of oxygen retained in the galaxy ISM is estimated with $M_{\rm Oxy}^{\rm ISM}$, and the fraction would still be a few percent even if $M_{\rm Oxy}^{\rm ISM}$ is underestimated by a factor of 2-3.
The large fraction of oxygen being removed by outflow would get even higher if $M_{\rm Oxy}^{\rm out}$ is underestimated.

\section{SED fitting results}\label{apx:sed_example}
\begin{figure*}[t]
\centering
\includegraphics[width=0.95\textwidth]{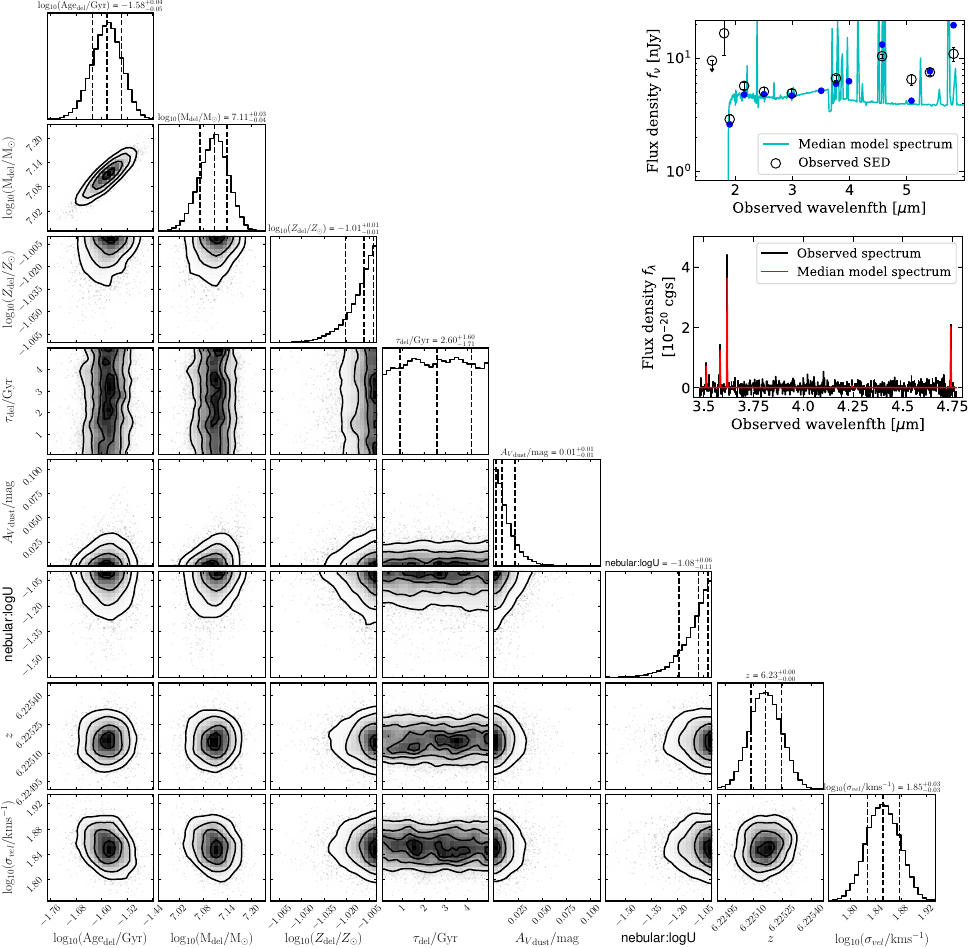}
\caption{SED fitting results for GLIMPSE-9081. The corner plot shows the posteriors of the eight free parameters (age, stellar mass, metallicity, $\tau_{\rm del}$, dust attenuation, ionization parameter, redshift, and emission line width). Note that the lens correction is not applied.
The observed SED and spectrum are shown on top right, along with the marginalized model spectrum.
}
\label{fig:SED_9081}
\end{figure*}

\begin{figure*}[t]
\centering
\includegraphics[width=0.95\textwidth]{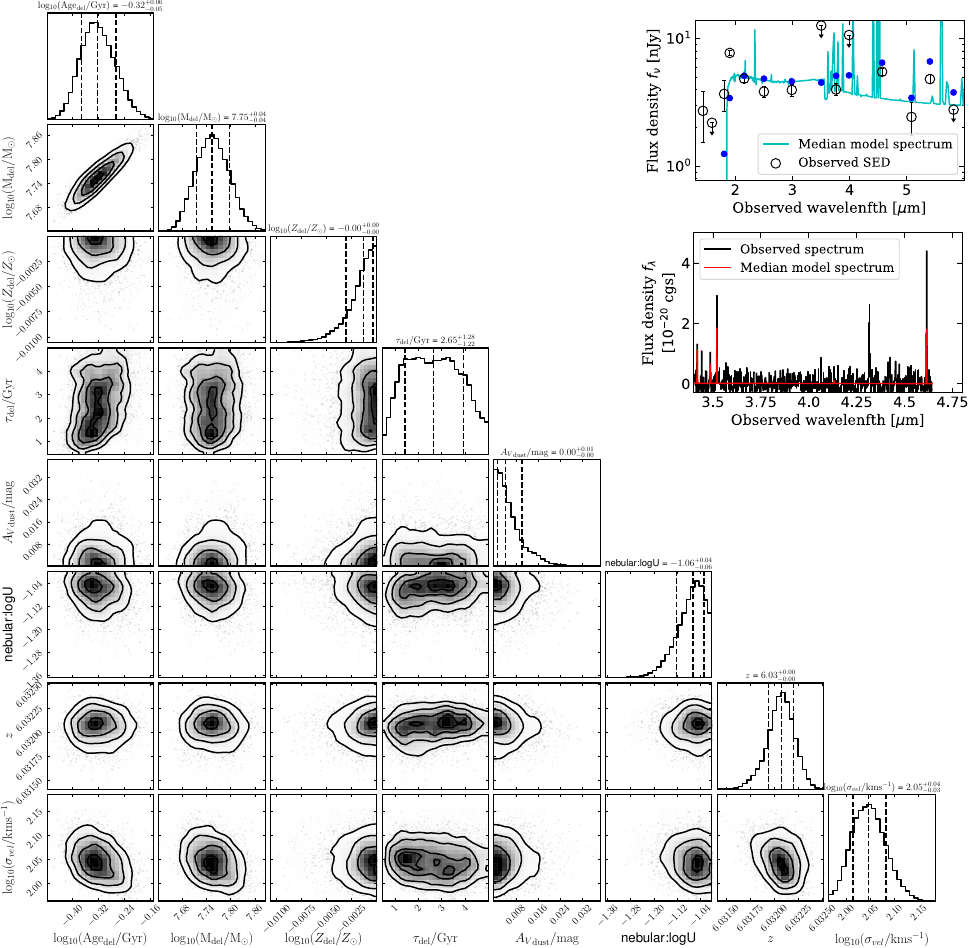}
\caption{Same as Figure~\ref{fig:SED_9081} but for GLIMPSE-17070.
}
\label{fig:SED_17070}
\end{figure*}

This section presents the spectro-photometry SED fitting results by \texttt{Bagpipes} obtained in Section \ref{subsec:prop}. The corner plots of the eight free parameters and the best-fit template spectra are displayed for two representative galaxies in the sample (GLIMPSE-9081 and GLIMPSE-17070).
These two sources are chosen because their NIRCam F150W S/Ns are the closest to the median of the sample (S/N$=11.3$ and $10.5$, respectively, and the median S/N is 10.8).
Therefore, Figure~\ref{fig:SED_9081} and \ref{fig:SED_17070} show the representative quality of our spectro-photometry fitting.

GLIMPSE-17070 is also one of the high-$\xi_{\rm ion}$ galaxies that exceeds the maximum $\xi_{\rm ion}$ values in BPASS SSP models.
As seen in Figure \ref{fig:SED_17070}, \texttt{Bagpipes} models cannot achieve such a high $\xi_{\rm ion}$ value, and the best-fit spectrum has a weaker H$\alpha$ line than the observation but gives a decent fit to the continuum levels and the overall SED shape.
We thus infer that the stellar mass estimates are rather robust even though the SSP library used in the \texttt{Bagpipes} fit are not able to fully replicate the emission line strength.

\bibliography{sample701}{}
\bibliographystyle{aasjournalv7}



\end{document}